\documentclass[sn-mathphys]{sn-jnl}

\jyear{2021}%

\numberwithin{equation}{section}

\def\epsilon{\varepsilon}

\usepackage{mathtools}
\DeclarePairedDelimiter\ceil{\ensuremath{\lceil}}{\ensuremath{\rceil}}
\DeclarePairedDelimiter\floor{\ensuremath{\lfloor}}{\ensuremath{\rfloor}}

\usepackage{makecell}

\newcommand{\Rn}{\mathcal{R}_0}
\newcommand{\Rt}{\mathcal{R}_t}
\newcommand{\ds}{\text{d}s}
\newcommand{\prob}{\mathbb{P}}
\newcommand{\dt}{\Delta t}

\begin{document}

\title[]{A hospital demand and capacity intervention approach for COVID-19 in the UK}

\author[1]{\fnm{James} \sur{Van Yperen}}

\author[2]{\fnm{Eduard} \sur{Campillo-Funollet}}

\author[1]{\fnm{Rebecca} \sur{Inkpen}}

\author[3]{\fnm{Anjum} \sur{Memon}}

\author*[1]{\fnm{Anotida} \sur{Madzvamuse}}\email{a.madzvamuse@sussex.ac.uk}

\affil*[1]{\orgdiv{School of Mathematical and Physical Sciences}, \orgaddress{\street{Department of Mathematics}, \orgname{University of Sussex}, \city{Brighton}, \postcode{BN1 9QH}, \country{UK}}}

\affil[2]{\orgdiv{School of Mathematical, Statistical and Actuarial Sciences}, \orgaddress{\street{Department of Statistical Methodology and Applications}, \orgname{University of Kent}, \city{Canterbury}, \postcode{CT2 7PE}, \country{UK}}}

\affil[3]{\orgdiv{Department of Primary Care and Public Health}, \orgaddress{\street{Brighton and Sussex Medical School}, \city{Brighton}, \postcode{BN1 9PH}, \country{UK}}}

\abstract{The mathematical interpretation of interventions for the mitigation of epidemics and pandemics in the literature often involves finding the optimal time to initiate an intervention and/or the use of infections to manage impact. Whilst these methods may work in theory, in order to implement they may require information which is likely not available whilst one is in the midst of an epidemic, or they may require impeccable data about infection levels in the community. In practice, testing and cases data is only as good as the policy of implementation and the compliance of the individuals, which means that understanding the levels of infections becomes difficult or complicated from the data that is provided. In this paper, we aim to develop a different approach to the mathematical modelling of interventions, not based on optimality, but based on demand and capacity of local authorities who have to deal with the epidemic on a day to day basis. In particular, we use data-driven modelling to calibrate an Susceptible Exposed Infectious Recovered-Died (SEIR-D) model to infer parameters that depict the dynamics of the epidemic in a region of the UK. We use the calibrated parameters for forecasting scenarios and understand, given a maximum capacity of hospital healthcare services, how the timing of interventions, severity of interventions, and conditions for the releasing of interventions affect the overall epidemic-picture.}

\keywords{COVID-19, mathematical, modelling, interventions, healthcare, data-driven}

\maketitle

\section{Introduction}

The resurgence of the severe acute respiratory syndrome coronavirus 2 (SARS-Cov-2) that causes COVID-19, and its mutant variants, are putting national health systems in most countries under significant pressure/strain due to an increase in COVID-19 hospitalisations and the provision of critical care for patients in need. Towards the end of 2020 and the beginning of 2021 the UK was battling the spike of COVID-19 infections across the country, which likely started in the Kent region, due to the spread of the Alpha variant. Since the end of 2021, the UK has been battling the spike of infections across the country, which this time likely started in London, due to the spread of the Omicron variant. The UK government has been pushing a combination of non-pharmaceutical interventions for England, such as the four step roadmap out of lockdown, which relies on the surveillance of the infection rate, and the effectiveness of the national vaccination programme to combat the spread of the Alpha variant. Although the success of previous lockdowns and tiered systems have come under scrutiny \cite{DBJal20}, the success of the roadmap interventions and the vaccination programme until approximately late May 2021 can be seen by examining the reported COVID-19 data on the Coronavirus Dashboard produced by the UK Government\footnote{\url{https://coronavirus.data.gov.uk/}}. From May onwards we start to see an increase in the number of cases nationally, which had not materialised into a significant increase in hospitalisations until the Omicron wave of December 2021. Indeed, in July 2021, the reported daily cases in the UK were around the same amount compared to the spike at the beginning of 2021 which, amongst other factors, caused the UK government to initiate ``lockdown 3", whereas daily hospitalisation counts in July 2021 were a quarter of what they were in January 2021. However, as we moved into winter, when infectious disease hospital activity is normally at its peak, with competing infections such as influenza, population healthcare management in hospitals and local authorities needed to be able to forecast the potential impact COVID-19 resurgence on hospital demand and capacity. Elective treatments, such as surgery and chemotherapy, have been and still are substantially backlogged due to the pandemic, and another resurgence of COVID-19 could add more pressure on healthcare systems without proper planning. Given the current situation, and the current state-of-the-art modelling of epidemics, hospital planners need to be able identify indicators in community infections which would cause a surge in hospital activity - in particular, using predictive modelling, they need metrics and rules as to when a potential surge may be occurring given the current data. This concept is particularly important with the reduction in the collection of COVID-19 data at the end of March 2022, with the implication that we are looking to ``live with" COVID-19 now, a.k.a the endemic phase, and so healthcare systems need quantitative approaches to be able to identify when an increase in hospitalisations implies a larger incoming wave. What this study aims to contribute to the epidemiological, public health and mathematical communities are ways in which local authorities can use national information in their own trained mathematical models, models which are fitted to the data of their region, and to run plausible scenarios based on effective measures to control their hospital demand and capacity. 

In the literature, there is a plethora of pre-prints and peer-reviewed articles that present work on forecasting COVID-19 dynamics, which can be used for future pandemics when they arise \cite{CVAal20,DZSal20,EHMal20,FLNal20,FGLal20,GLBal20,JLJ20,LTY20,LMKal20,MG20}. With the rise of data-science over the last decade, data availability and quality has enabled data-driven modelling and research, allowing for clear applications of infectious disease modelling rather than just theoretical work. In particular, the availability of government-led data initiatives making epidemiological and public health data accessible \cite{Ar20}. For example, here in the UK, during the COVID-19 pandemic the government produced the Coronavirus Dashboard which gives public access to local, regional and national data concerning testing, cases, hospitalisations, vaccinations and deaths - all major contributions when considering a data-driven model of healthcare demand and capacity. Modelling efforts now should be conducted in such a way which allow them to be used in the future, to which we can use COVID-19 data and policy as an application of the work. However, the majority of mathematical modelling publications are aimed either at national level modelling with an assumption that the reader knows the standard mathematical jargon. The inherent assumptions behind the modelling decisions and the parameters that are adjusted for interventions scenarios are often not made clear. The swift wave of COVID-19 across the globe has identified the need for reliable, sensitive and validated data-driven approaches that are accessible by local authorities to make quantitative and qualitative decisions on policy. To combat this, public-policy in mind, epidemiological research groups across the UK, and in fact across the world, have been producing web-based tools to combat COVID-19 and provide ways for non-mathematicians to picture and understand the data available. A comprehensive review of different web-based tools can be found in \cite{MHG21}. Since these models are readily available to be used, and with the conclusion and recommendations of the Goldacre report for public healthcare management to ``embrace help from other sections such as academia" \cite{goldacre}, it is more important now more than ever that the mathematically modelling assumptions are present, visible and understandable and that the scope of the model is clear, something which we look to obtain with our toolkit Halogen\footnote{\url{https://www.halogen-health.org/}}. 

Although we focus on COVID-19 and use COVID-19 data to fit our model, we note that the concept behind this study can be applied to any infectious disease, or other application, provided it has an SIR type model describing the mathematics and there is the appropriate data available. Whilst it is reasonably simple to generate an SIR type model for an infectious disease, COVID-19 is seemingly the first infectious disease to have data collected and made publicly available for parameter estimation in the manner we present, and so we can not claim that our approach would work without such detailed data. One may assume that, in the scenario of future pandemics, similar such government-led data initiatives will be available to enable future modelling efforts. Indeed, similar conceptual approaches have been used by modelling groups across the country to understand the impact of the UK Foot and Mouth outbreak of 2001, see \cite{K05} and references therein, and the UK SARS H1N1 pandemic of 2009, see \cite{BVJFWE10} and references therein. For a mathematical modelling approaches to a generic influenza pandemic at a national or international level, see also \cite{FCFCCB06} and references therein. 
 
The objective of this paper is to demonstrate the flexibility of data-driven epidemiological and mathematical modelling for providing robust intervention scenarios. In this study we focus on hospital capacity as the metric to decide if an intervention is initiated. We consider the situation whereby given information about the spread of the disease, in particular the ``known" $R$ number, we demonstrate how using current hospital capacity to lift an intervention can significantly change the epidemic and demonstrate an approach to judging what the thresholds of occupied beds should be given a resource limit (maximum number of beds available for COVID-19 patients) before initiating an intervention. Similar studies have been conducted using ICU capacity as a determinant, such as \cite{AFG22,MSW20}, best intervention scenario possible given parameters. such as \cite{DKM21,NPP16}, or only consider intervention scenarios based on the current number of infections, such as \cite{FIL21,JG22,S21}. What we aim to demonstrate in this paper is the ability to conduct decision making that does not require information not available in the moment (such as the peak of epidemic curves).

The outline of this paper is as follows. In Section \ref{sec:seird} we demonstrate the flexibility of our approach in \cite{CVAal20} by fitting parameters of regions with publicly available data. In Section \ref{sec:EBM} we apply intervention measures based on hospital occupancy using the parameters previously inferred. In Section \ref{sec:ABM} we introduce an agent-based equivalent to the mathematical model in Section \ref{sec:seird}, validate the approach and apply the intervention scenarios in Section \ref{sec:EBM}, conduct a further investigation into how transmission and contacts effect the results of the scenarios and compare the results to the results in Section \ref{sec:EBM} to understand the role of stochastic perturbations. In Section \ref{sec:further} we outline some limitations of this work with potential solutions and in Section \ref{sec:conc} we summarise the main findings of this study.

\section{A data-driven SEIR-D model}
\label{sec:seird} 

\subsection{Data}

\begin{figure}[!hbt]
    \centering
    \includegraphics[width=\linewidth]{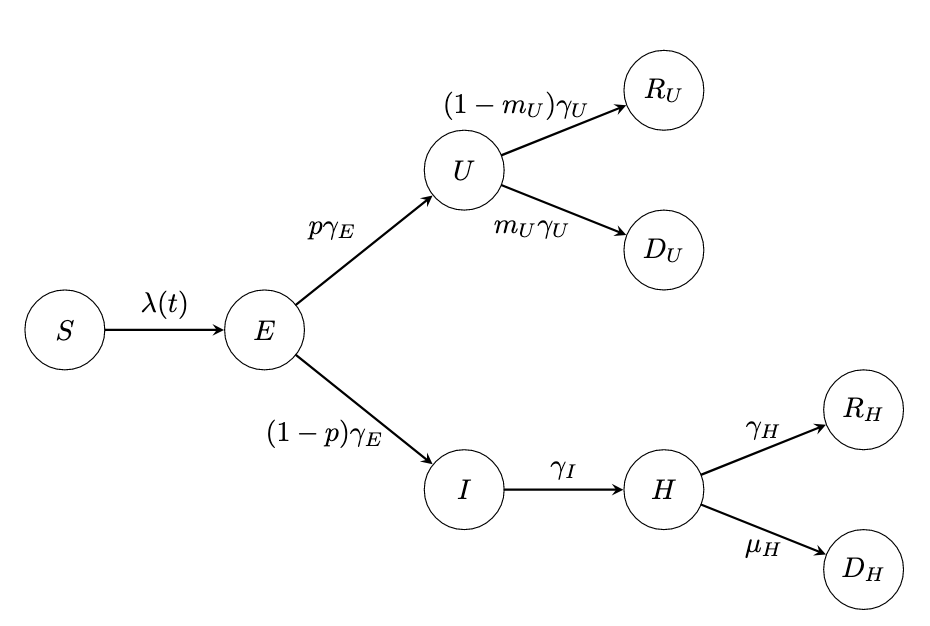}
    \caption{Schematic representation of the compartmental pathways on which the SEIR-D mathematical model is formulated used first principles.}
    \label{fig:model}
\end{figure}

Following \cite{CVAal20}, in this section we present a simple data-driven susceptible-exposed-infected-recovered-dead (SEIR-D) model as demonstrated in Figure \ref{fig:model}. The model breaks down the typical infectious compartment of an SEIR-D model into two compartments, one strand to model individuals who become infectious with COVID-19 and will be going to hospital ($I$), and the other strand to model individuals who will not need to go to hospital and thus remain undetected by hospital healthcare requirements ($U$). Splitting the infectious compartment this way allows us to describe the number of individuals who are in hospital with COVID-19 at any time, which allows us to fit the hospital data to the model to be able to infer information about the number of infectious cases in the community. This approach allows us to circumvent the issue of parameter identifiability and estimation due to the lack of obtainable and usable data regarding those who are infectious and asymptomatic. 

In \cite{CVAal20} we used specific regional datasets to calibrate the model. As part of the national COVID response, all the National Health Service (NHS) hospitals in England treating COVID-19 patients submitted a Daily Situation Report (SITREP) to NHS England. The data associated to Sussex NHS Trusts were extracted and combined to weekly death counts from the Office for National Statistics (ONS) with COVID-19 reported as the underlying cause of death, which we then received and fitted. To be precise, by Sussex we mean the collective term for geographies pertaining to the counties of East Sussex and West Sussex in South East England. The SITREP contained counts of daily admissions, daily discharges, and the beds occupied daily, whilst the death dataset contained the number of deaths recorded outside of hospitals and the number of deaths recorded within hospitals. 

Whilst the ONS death data are publicly available, the SITREP data in general are not, but, given the national need for data, the UK government produced the Coronavirus Dashboard. It provides users with the ability to look at different metrics of COVID-19, such as hospitalisations and deaths, for different regions and provides an API for users to download the datasets themselves. The granularity of the data depends on the size of the region the data are required for, typically all data are available for each nation of the UK, but hospitalisations and deaths are split depending on their geographical location. Indeed, hospitalisations are recorded using NHS regions and NHS trusts, and deaths are recorded using local authority regions. The unfortunate difference between using the Coronavirus dashboard and the SITREP is that the Coronavirus dashboard does not contain the number of daily discharges, and it does not differentiate place of death like the ONS weekly death registry does. In order to apply the approach presented in \cite{CVAal20} we adapted the Coronavirus dashboard data in the following way: we calculated the proportion of hospital COVID-19 deaths to the total number of COVID-19 deaths for the region being considered from the ONS weekly death registry and applied this proportion to the deaths dataset acquired from the Coronavirus Dashboard to give us a proxy dataset on deaths in hospital and deaths outside of hospital. Next, we then used the deaths in hospital with the admissions and beds occupied to find a proxy dataset for the discharges, since the offset of beds occupied between each day depends on the number of admissions, discharges and deaths that day. We note that, although the proxy death datasets might closely track the ONS weekly death registry, the discharges dataset is very noisy due to noise accumulating from multiple sources. For this reason, we decided to apply a 7-day rolling average to the discharges. We note here that the Coronavirus dashboard has access to patients in mechanical ventilation beds and so we could include a compartment that describes the high-dependency unit (HDU). We decided not to do this since we do not have access to the number of patients who have died in HDU, we only have access to the number of patients who died in hospital, and so parameter identifiability would be an issue. 

For illustrative purposes, we only consider the geographical regions of North West England, South East England and the nation of England. We chose these as both NHS region and local authority region are the same.

\subsection{Model and parameter estimation}

The mathematical model is a data-driven generalisation of a simple temporal epidemiological dynamic system of ordinary differential equations, governed by the interactions depicted in Figure \ref{fig:model}, supported by non-negative initial conditions
\begin{align}
    \dot{S} &= - \beta \frac{U + I}{N}S, \qquad \quad \, \, \, \quad t \in (0,T], \qquad S(0) = S_0, \label{Seq} \\
    \dot{E} &= \beta \frac{U + I}{N}S - \gamma_E E, \quad \quad t \in (0,T], \qquad E(0) = E_0, \label{Eeq} \\
    \dot{U} &= p \, \gamma_E E - \gamma_U U, \qquad \quad \, \, \, \, t \in (0,T], \qquad U(0) = U_0, \label{Ueq} \\
    \dot{I} &= (1 - p) \gamma_E E - \gamma_I I, \quad \, \, \, \, t \in (0,T], \qquad I(0) = I_0, \label{Ieq} \\
    \dot{H} &= \gamma_I I - (\gamma_H + \mu_H) H, \quad \, t \in (0,T], \qquad H(0) = H_0 \label{Heq} \\
    \dot{R}_U &= (1 - m_U) \gamma_U U, \qquad \quad \, \, \, t \in (0,T], \qquad R_U(0) = R_{U,0}, \label{RUeq} \\
    \dot{R}_H &= \gamma_H H, \qquad \qquad \qquad \quad \, \, t \in (0,T], \qquad R_H(0) = R_{H,0}, \label{RHeq} \\
    \dot{D}_U &= m_U \gamma_U U, \qquad \qquad \qquad t \in (0,T], \qquad D_U(0) = D_{U,0}, \label{DUeq} \\
    \dot{D}_H &= \mu_H H, \qquad \qquad \qquad \quad \, t \in (0,T], \qquad D_H(0) = D_{H,0}. \label{DHeq} 
\end{align}
Here, the dot above the notation denotes the time derivative. In this setting, $N$ denotes the total regional population. $S(t)$ denotes the proportion of the total population $N$ who are susceptible to the disease, COVID-19. Susceptible individuals become exposed to the disease, i.e. they are carrying the disease but are not currently infectious, to form the $E(t)$ subpopulation at rate $\lambda(t)$ which represents the average infectivity. The rate $\lambda(t)$ is the product between $\beta$, the average transmission rate, and the probability of meeting an infectious person $(U(t) + I(t))N^{-1}$. The $E(t)$ subpopulation is an incubation compartment and further evolves in two ways. A proportion of $E(t)$ becomes infectious but, in the spirit of the hospital healthcare demand, remains undetected with probability $p$, denoted $U(t)$, at a rate $\gamma_E$, or a proportion of $E(t)$ becomes infectious and will require hospitalisation in the future with a probability of $1-p$, denoted $I(t)$, at a rate $\gamma_E$. The subpopulation that does not require hospitalisation can either progress to recover with a probability of $1-m_U$, at rate $\gamma_U$, to form the recovered population, denoted by $R_U(t)$, or die with a probability of $m_U$, at rate $\gamma_U$, to form the dead population, denoted by $D_U(t)$. Considering the infectious population that will be going to hospital, these individuals will become hospitalised, denoted by $H(t)$, and thus be in hospital care at rate $\gamma_I$. We assume that once a patient has been admitted into hospital, they no longer transmit to other non-COVID-19 patients, visitors or workers. This could be due to appropriate isolation and use of personal hygiene measures in hospital, however we do note that transmission in hospital is not unheard of. For the model, this means that we assume transmission is only conducted in the community. Once in hospital, patients can evolve in two separate pathways, a proportion of the hospitalised population can fully recover at rate $\gamma_H$ to form the subpopulation denoted by $R_H(t)$. Alternatively, if they can not recover, then they die while in hospital at rate $\mu_H$, to form the dead population denoted by $D_H(t)$. We note that $\gamma_H$ and $\mu_H$ are closely related to the length of stay in hospitals. 

As is standard for epidemiological models of this nature, $\beta$ denotes the average transmission rate, $\gamma_E^{-1}$ denotes the average incubation time, $p$ denotes the average proportion of infectious individuals who will not require hospital treatment, $\gamma_U^{-1}$ denotes the average infectious period for those not needing hospital treatment, $m_U$ denotes the infected fatality ratio for undetected cases, $\gamma_I^{-1}$ denotes the average infectious period from becoming infectious to being admitted to hospital, $\gamma_H^{-1}$ denotes the average hospitalisation period for those who recover and $\mu_H^{-1}$ represents the average hospitalisation period for those who die. For this model, using the method of next generation matrices \cite{DHR10}, we derive the formula for the basic reproduction number $\Rn$ as the following
\begin{align}
    \label{R0calc}
    \Rn := \beta \left( \frac{p}{\gamma_U} + \frac{1-p}{\gamma_I} \right). 
\end{align}
Using this, we calculate the effective reproduction number $\Rt$ as
\begin{align}
    \label{Rtcalc}
    \Rt := \Rn \, \frac{S(t)}{N}.
\end{align}
The effective reproduction number $\Rt$ is often referred to as the $R$ number. We briefly explain the fitting procedure presented in \cite{CVAal20}, we refer the interested reader to \cite{CVAal20} for further details. First, we utilise the linear relationship between the model description of hospital discharges and hospital deaths and use linear regression analysis to calculate the ratio of discharges to deaths. One can see that, in terms of the model and its parameters, the daily discharges can be written as
\begin{align}
    Dis(t) := \gamma_H \int_{t-1}^t H(s) \, \ds, \label{dis}
\end{align}
and the daily hospital deaths can be written as
\begin{align}
    Dth_H(t) := \mu_H \int_{t-1}^t H(s) \, \ds. \label{dhp}
\end{align}
The linear regression allows us to estimate the ratio between $\gamma_H$ and $\mu_H$. Next, we rewrite equations \eqref{Seq}--\eqref{Heq} in terms of the data, we call this the ``observational" model, whereby \eqref{RUeq}--\eqref{DHeq} are not considered since they are cumulative representations of the compartments in \eqref{Seq}--\eqref{Heq}. The observational model is formed of compartments that are available from the data. Indeed, one can see that, in terms of the model and its parameters, the daily admissions can be written as
\begin{align}
    Adm(t) := \gamma_I \int_{t-1}^t I(s) \, \ds, \label{adm}
\end{align}
the daily deaths outside of hospital is
\begin{align}
    Dth_U(t) := \gamma_U m_U \int_{t-1}^t U(s) \, \ds, \label{dnh}
\end{align}
and daily discharges as in \eqref{dis} above. Thus, the observational model, as presented in \cite{CVAal20}, is
\begin{align}
    \dot{H} &= \gamma_I I - \gamma_H\left(1 + \frac{\mu_H}{\gamma_H}\right) H, \label{obs1} \\
    \dot{U} &= \frac{p}{1-p} \left( \dot{I} + \gamma_I I \right) - \gamma_U U, \label{obs2} \\
    \dddot{I} &= \left(\ddot{I} + (\gamma_E + \gamma_I) \dot{I} + \gamma_E\gamma_I I \right)\left[ \frac{p \gamma_I I}{1 - p} + \frac{\dot{I}}{1-p} - \gamma_U U - \frac{\beta}{N} (U+I)^2 \right](U+I)^{-1} \nonumber \\
    & \qquad - (\gamma_E + \gamma_I) \ddot{I} - \gamma_E\gamma_I \dot{I}. \label{obs3} 
\end{align}
We solve the observational model, compute \eqref{dis}, \eqref{adm} and \eqref{dnh} and compare against the datasets given. This allows us to use a maximum likelihood estimation approach by minimising a negative log-likelihood with some constraints on the initial conditions and on the effective reproduction number. These constraints are needed to keep the feasible region of the parameters close to the realistic sets. Without these constraints, the peak of the datasets could be explained by being close to herd immunity (i.e., a lot of infections have already occurred before the lockdown) or a large value of $\Rt$ combined with a small value of $p$. We note here that \eqref{dhp} does not need to be used in the observational model due to the linear regression and the fact that the resulting log-likelihood functions from the regression and observational model are independent. This means that there is one less parameter to infer from solving the observational model.

Since we are considering the data from the first day of lockdown, we also need to infer the initial conditions. We have not conducted a formal investigation into the resulting log-likelihood, but it is clear that there is a continuous dependence between initial conditions and the parameters, see \cite{CWVDM21} for a comprehensive discussion. In practice, we see this manifest as an issue to fit $p$ - if the first guess of initial conditions and parameters is not close to the ``true" values, then it is $p$ which changes in value the most. Although in reality, $p$ is characterised by how COVID-19 affects individuals, from demographics such as age, ethnicity, gender, however we speculate this value should not change drastically between regions. In view of this, in the parameter estimation process we fix $p$ to be the same throughout the estimation for the three regions. 

\begin{figure}[!ht]
    \centering
    \includegraphics[width=0.75\linewidth]{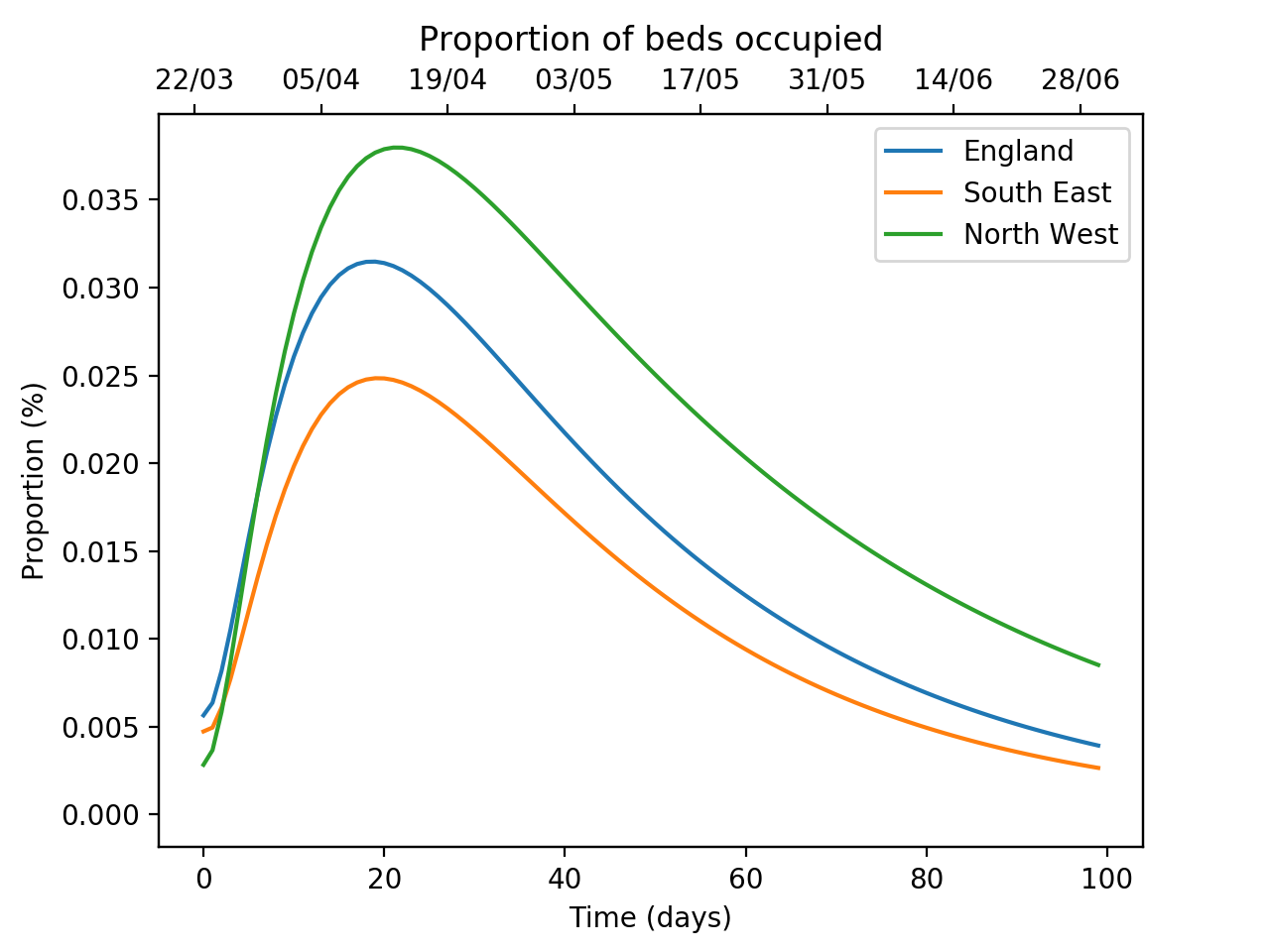}
    \caption{Proportional of beds occupied for England, South East and North West.}
    \label{fig:prop_beds}
\end{figure}

\begin{figure}[!ht]
    \centering
    \includegraphics[width=0.9\linewidth]{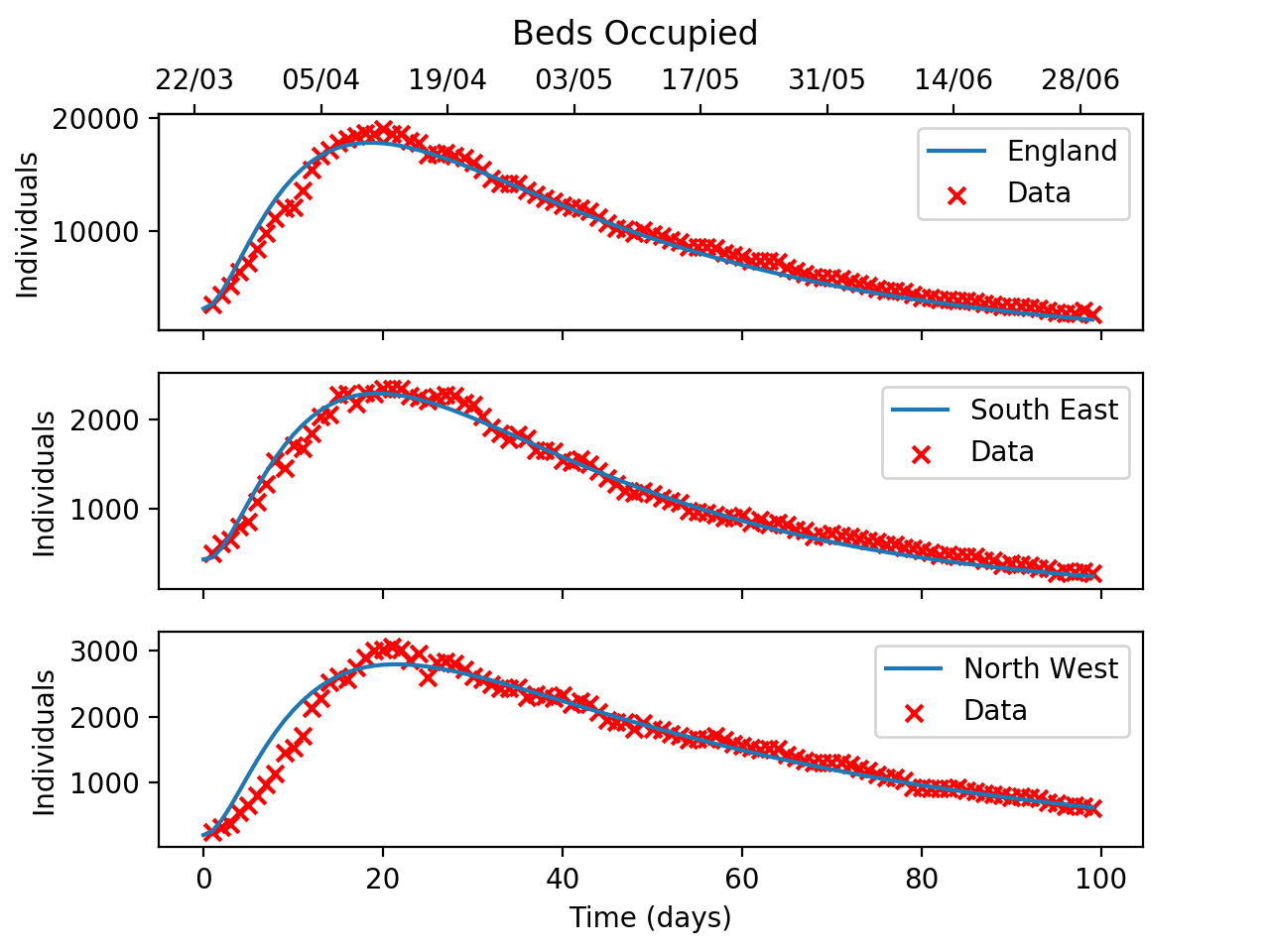}
    \caption{Beds occupied fit for England, South East and North West.}
    \label{fig:beds_occupied}
\end{figure}

From the model and using the fitting procedure described above, we gain the parameters in Table \ref{tab:all_params} and initial conditions in Table \ref{tab:all_ic}, with a demonstration of the fit for beds occupied demonstrated in Figure \ref{fig:beds_occupied}. The equivalent figure for the Sussex region fit can be found in \cite{CVAal20} in Figure 2. We note that the infected fatality ratio and average hospitalisation period for those who recover are similar across all regions, but the average hospitalisation period for those who die and the value of $\Rt$ when lockdown commenced are quite different. The varied values of $\Rt$ could be explained by the amount of infections in each region, it was reported at the time that Sussex and the South East escaped quite lightly on the number of infections, whilst the North of England did not. This is shown in Figure \ref{fig:prop_beds}, whereby we have taken the model output of the beds occupied and divided it by the population size for each region and multiplied it by 100. Proportionally, the North West saw almost double the amount of patients as the South East. The higher value of $\Rt$ for North West can also be seen by looking at the gradient of the beds occupied. There could be a myriad of reasons for this, such as the demographic of the population or the geography of the region. One should also note that, as we came out of the first lockdown and went into the first iteration of the tiered system, it was cities in the north which first started to show signs of resurgence (such as Manchester). Using $\gamma_H$ and $\mu_H$ for each region, one can calculate that the estimate of the probability of discharge for England, South East and North West was 58.03\%, 61.57\%, 55.31\% respectively. As for the reason why, one may speculate that this is due to the number of patients in the hospitals being higher in the north which puts pressure on the health system.

\begin{table}[ht!]
    \centering
    \caption{Parameters of interest derived using method in \cite{CVAal20} for three different regions of the UK.}
    \label{tab:all_params}
    \begin{tabular}{l l l l}
        \toprule
        Parameter & England & South East & North West \\
        \midrule
        $\Rt$ & 0.763 & 0.733 & 0.835 \\ 
        $\gamma_H^{-1}$ & 13.11 days & 13.20 days & 13.28 days \\
        $m_U$ & 0.0013 & 0.0013 & 0.0010 \\ 
        $\mu_H^{-1}$ & 18.13 days & 21.16 days & 16.43 days \\
        \botrule
    \end{tabular}

\end{table}   

\begin{table}[ht!]
    \centering
    \caption{Initial conditions of interest derived using method in \cite{CVAal20} for three different regions of the UK.}
    \label{tab:all_ic}
    \begin{tabular}{l l l l}
        \toprule
        Initial Condition (\%$N$) & England & South East & North West \\
        \midrule
        $E_0$ & 0.891\% & 0.710\% & 1.061\% \\ 
        $U_0$ & 0.025\% & 0.018\% & 0.025\% \\ 
        $I_0$ & 0.041\% & 0.001\% & 0.002\% \\
        \botrule
    \end{tabular}
\end{table}  

\section{Intervention measures based on hospital occupancy}
\label{sec:EBM}

For the remainder of this study we will use the parameters identified in Tables \ref{tab:all_params} but we will fix the initial conditions to represent the beginning of the epidemic. Namely, in \eqref{Seq} and \eqref{Eeq} we set
\begin{align}
    S_0 = \ceil*{0.999N}, \quad \text{and} \quad E_0 = \floor*{0.001N}, \label{initial_conds}
\end{align}
and for the remaining equations \eqref{Ueq}--\eqref{DHeq} we set the initial conditions equal to 0, where $\ceil*{x}$ represents the ceiling of $x$ and $\floor*{x}$ represents the floor of $x$. 

Using \eqref{Seq}--\eqref{DHeq} we will model an intervention as a social distancing effect, by manipulating the average transmission rate $\beta$. Throughout this study we will be measuring the ``success" of an intervention by the percentage of individuals who have died throughout the simulation, in the sense that reducing this statistic means a more successful intervention. 

The outbreak is regarded to have been contained at a time $T>1$ such that $E(T) + U(T) + I(T) < 1$, $\mathcal{R}_t < 1$ and there is no ongoing intervention. This implies that herd immunity has been achieved and thus the system has reached its steady state. This description highlights one of the drawbacks of using continuum equations over their stochastic counterparts, namely in the continuum setting there is always ``some amount" of the disease leftover in the community (i.e. $0 < E(t) + I(t) + U(t)$). This can amount to another outbreak if the parameters are changed appropriately, which we would not expect to happen in reality. 

We numerically approximate the solutions to the system \eqref{Seq}--\eqref{DHeq} by using the SciPy implementation of the ``lsoda" method, which is a combination of the Adams methods and the Backward Differentiation Formula (BDF) family of methods \cite{BG08,GH10,P83}. Given the multi-step approach of the ODE solver, each time we manipulate the parameters during a simulation, to initiate an intervention, we stop the current solver and start it again using initial conditions as the final values of the last solver. This bypasses difficulties of having a discontinuous ODE system (with respect to the parameters). One notes that this can also be bypassed by using a much simpler solver, like the forward Euler scheme, however this would result in the need for significantly smaller timesteps.

\subsection{The ``do-nothing" approach}

\begin{figure}[!htb]
    \centering
    \includegraphics[width=0.6\linewidth]{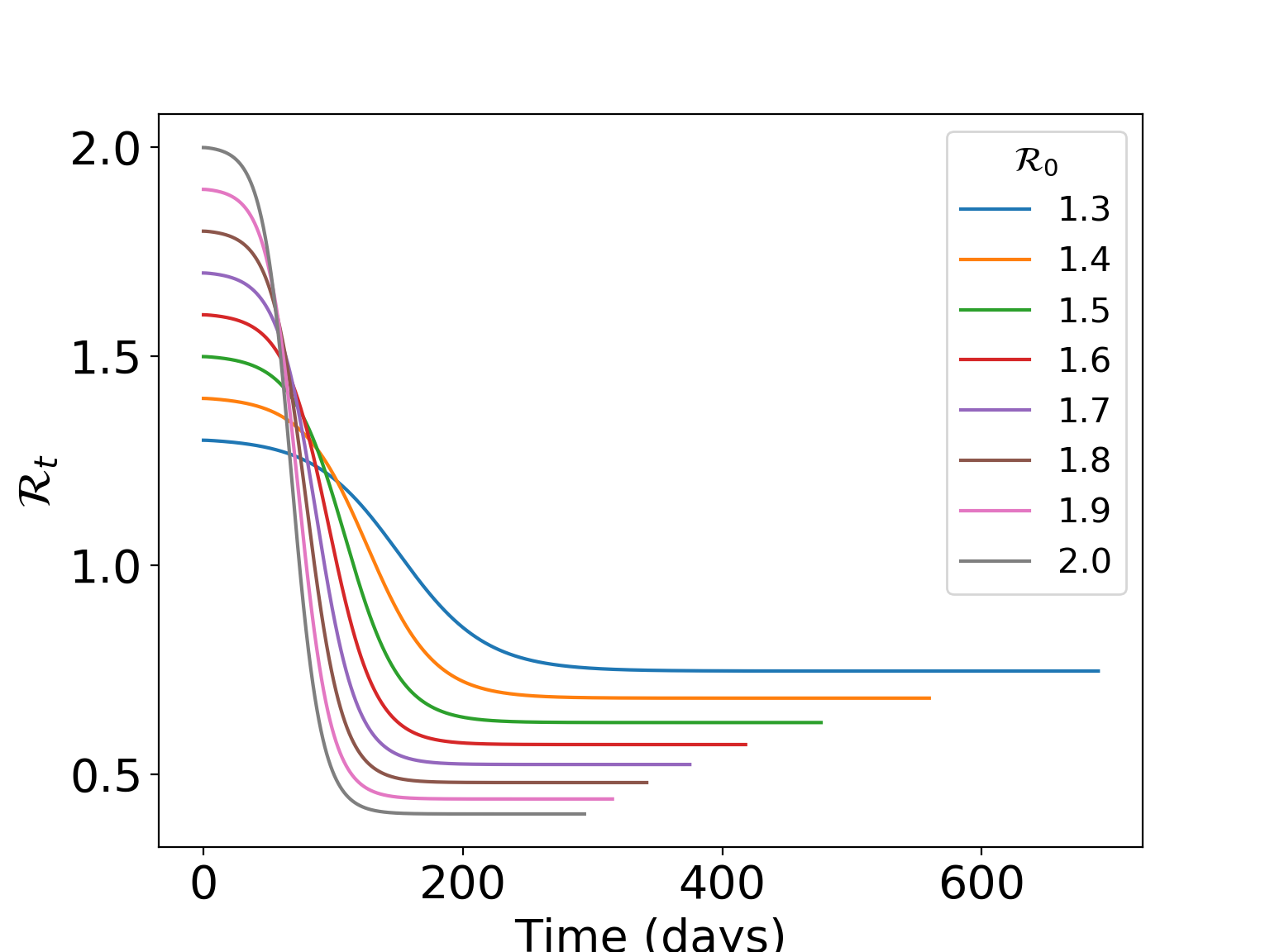}
    \caption{$\Rt$ using the ``do-nothing" approach with the England parameters. Here North West is abbreviated to NW and South East is abbreviated to SE.}
    \label{fig:donoRt}
\end{figure}

\begin{figure}[!htb]
    \centering
    \includegraphics[width=0.85\linewidth]{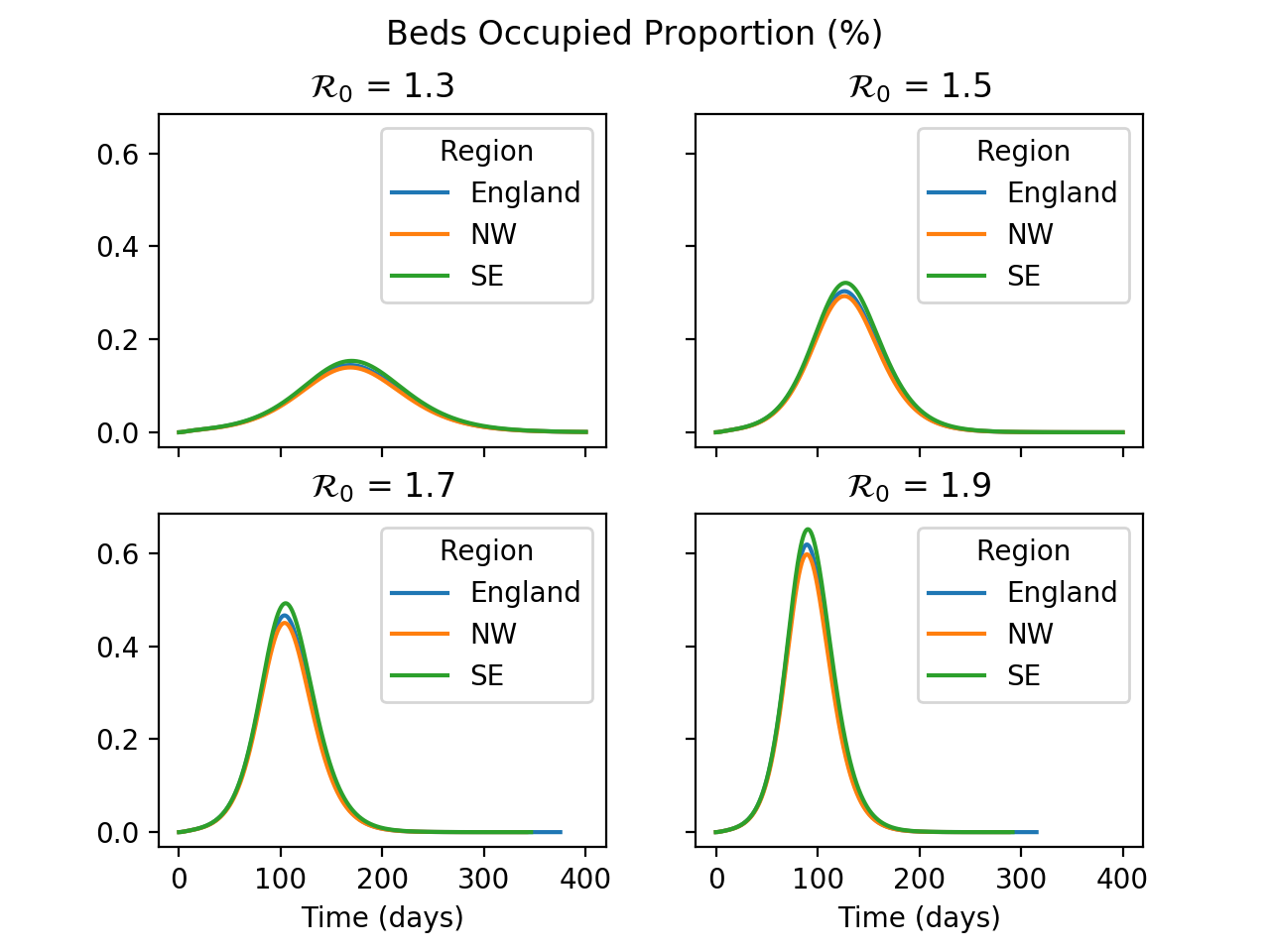}
    \caption{Percentage of beds occupied each day using the ``do-nothing" approach for each region. We note that we have truncated the simulation to make visualisation easier. Here North West is abbreviated to NW and South East is abbreviated to SE.}
    \label{fig:donoH}
\end{figure}

As a reference to compare how the interventions are working, we use this section to demonstrate the ``do-nothing" approach, which is simply to let the disease take its course. This will provide us with statistics to compare to the interventions later on to demonstrate their effectiveness. In reality, we are aware that this approach will not be implored in practice and an intervention will occur, as has happened all over the world with national level lockdowns and social distancing measures. Since the parameters presented in Table \ref{tab:all_params} depict a lockdown scenario, we scale $\beta$ appropriately to several values to establish an epidemic, i.e. so that $\Rn > 1$. An increase in average transmission rate can simply be interpreted as more individuals meeting each other and spreading the disease. In Table \ref{tab:dono} we measure the maximum hospital capacity needed as a percentage of the total population, the day in the simulation that maximum is reached and the percentage of dead individuals at the end of the outbreak, for each region. As intuitively expected, as $\Rn$ increases the percentage of maximum number of patients in the hospital increases, the day of that peak is sooner and the percentage of dead individuals increases. Interestingly, the day of the peak does not seem to change even though the hospitalisation parameters are all quite varied, we suspect that this is somehow related to the initial conditions and the fact that $\mathcal{R}_0$ is not effected by the parameters that describe hospitalisations. We see that the percentage of maximum beds occupied is worse in the South East, followed by England and then the North West respectively. This is to be expected since the average hospitalisation periods (for deaths) are longer in the respective order which implies more people in hospital at any one moment. However, the percentage of dead individuals is higher in the North West, followed by England and then the South East respectively. This is also to be expected since the probability of death in hospitals is larger in the respective order whilst the mortality probability outside of hospitals are very similar for each of the regions. In Figure \ref{fig:donoRt} we demonstrate the effective reproduction number $\Rt$ of the simulation using the England parameters. This shows us that, when $\Rn$ is larger, the actual outbreak is much shorter in length and reaches much smaller values of $\Rt$. This description follows the notion that the larger the value of $\Rn$, the more aggressive the disease is following the exponential growth of those who are infectious, as can be seen in Figure \ref{fig:donoRt} by the steep decline in $\Rt$. In Figure \ref{fig:donoH} we demonstrate a comparison of the percentage of beds occupied for each of the regions and for some values of $\Rn$. We note that in Figure \ref{fig:donoH} we truncate the simulation to make the visualisation of the hospitalisations easier. 

\begin{sidewaystable}
\sidewaystablefn%
\begin{center}
\begin{minipage}{\textheight}
\caption{Measurements using the ``do-nothing" approach.}
\label{tab:dono}
\begin{tabular*}{\textheight}{@{\extracolsep{\fill}}lccccccccc@{\extracolsep{\fill}}}
\toprule%
& \multicolumn{3}{@{}c@{}}{Max beds occupied (\%$N$)}& \multicolumn{3}{@{}c@{}}{Peak of beds occupied (day)} & \multicolumn{3}{ c }{Dead individuals (\%$N$)} \\
\cmidrule{2-4}\cmidrule{5-7} \cmidrule{8-10}%
$\Rn$ & England & South East & North West & England & South East & North West & England & South East & North West \\
\midrule
1.3 & 0.145\% & 0.154\% & 0.140\% & 168 & 170 & 168 & 1.122\% & 1.032\% & 1.178\% \\ 
        1.4 & 0.222\% & 0.236\% & 0.214\% & 144 & 145 & 144 & 1.352\% & 1.244\% & 1.420\% \\
        1.5 & 0.304\% & 0.322\% & 0.293\% & 126 & 128 & 126 & 1.540\% & 1.417\% & 1.618\% \\
        1.6 & 0.386\% & 0.408\% & 0.372\% & 114 & 115 & 114 & 1.696\% & 1.560\% & 1.781\% \\
        1.7 & 0.466\% & 0.492\% & 0.450\% & 104 & 105 & 104 & 1.825\% & 1.679\% & 1.917\% \\ 
        1.8 & 0.544\% & 0.574\% & 0.525\% & 96 & 97 & 96 & 1.933\% & 1.779\% & 2.031\% \\ 
        1.9 & 0.619\% & 0.652\% & 0.598\% & 89 & 90 & 89 & 2.025\% & 1.863\% & 2.127\% \\ 
        2.0 & 0.686\% & 0.725\% & 0.666\% & 84 & 85 & 84 & 2.103\% & 1.935\% & 2.209\% \\
\botrule
\end{tabular*}
\end{minipage}
\end{center}
\end{sidewaystable}

\subsection{Notation for an intervention}
\label{intervention}

\begin{figure}[!htb]
    \centering
    \includegraphics[width=0.85\linewidth]{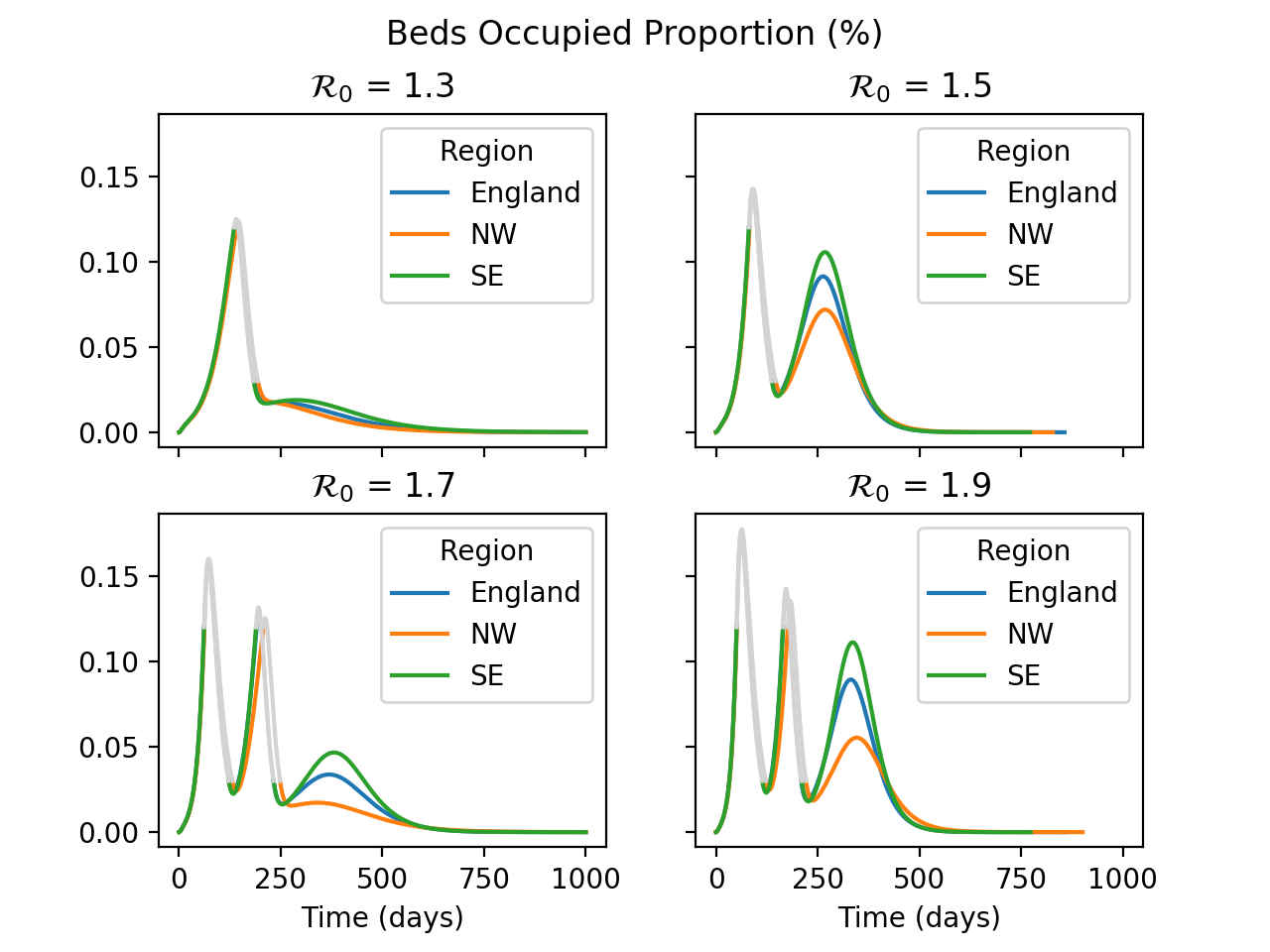}
    \caption{Percentage of beds occupied each day using the hospital capacity intervention scenario for each region. We fix $H_l := 0.25 H_u$. The grey line represents the time in an intervention. We note that we have truncated the simulation to make visualisation easier. Here North West is abbreviated to NW and South East is abbreviated to SE.}
    \label{fig:int_all_regions}
\end{figure}

In the following sections we want to investigate how one can use hospital capacity as a measurement for whether interventions are put into place. We aim to model the situation where an intervention is triggered when hospital capacity is almost full, and then finish the intervention when the hospital capacity has reached an ``opening" threshold of significantly lower patients. For ease of demonstration we will simply set a threshold that once breached will trigger the intervention, in some cases this will mean that the capacity is ``breached" - we deal with this in subsection \ref{minim_cap}. We denote the state of being in an intervention using the notation $\ell$, namely if $\ell=1$ then we are in an intervention, otherwise we set $\ell=0$. Using this, we describe the average transmission rate as 
\begin{align}
    \overline{\beta}(t;\ell) := [\ell = 1] \, \beta + [\ell = 0] \, C_{\Rn} \beta, \label{bbar}
\end{align}
where $\beta$ is the average transmission rate associated to the first lockdown deduced from the parameter inference for each region, $C_{\Rn}$ is a scaling constant to give the initial value of $\Rn$ wanted, and $[\cdot]$ are the Iverson brackets \cite{K92} defined as
\begin{align*}
    [P] := \begin{cases}
    1 & \text{ if } P \text{ is true,} \\
    0 & \text{ otherwise.}
    \end{cases} 
\end{align*} 
We describe the intervention in the following recursive way
\begin{align}
    \ell := [\ell = 0][H(t) > H_u] + [\ell = 1][H(t) > H_l], \label{ell}
\end{align}
which is to say that the intervention is triggered at time $t$ when the number of patients in hospital goes above an upper limit $H_u$, and the intervention stays triggered until the number of patients in hospital goes below a lower limit $H_l$. Initially $\ell$ is set equal to 0. The values of $H_u$ and $H_l$ are regionally dependent since they depend on the maximum capacity of all the hospitals in a region. Given the maximum hospital capacities in Table \ref{tab:dono}, we fix $H_u := 0.0012N$, to guarantee at least one lockdown for each $\Rn$ chosen, and vary the opening up threshold between $H_l := 0.025H_u$ and $H_l := 0.5H_u$. We will present the results from all the regions in various tables, however graphically we will restrict the figures to the South East region so the figures are not a visual burden. To this end, we present Figure \ref{fig:int_all_regions} which is the only comparison of the interventions for each region. The simulations all look reasonably similar and follow a similar trend, the noticeable differences come from when one region goes into an intervention (light grey coloured line) whilst the others don't. This explains some of the interesting numbers in Tables \ref{tab:R0_intervention}--\ref{tab:Hl_intervention}. We go into more detail in the following sections.  

There are some caveats to this study we want to highlight. It is somewhat unrealistic that transmission would immediately revert to normal amounts after an intervention \cite{MB20}, however we do not consider this here. Another aspect to consider with this study is that we are considering the capacity of all the hospitals in a region, due to modelling constraints and data access. This means that we are assuming hospitals can move patients throughout each region due the physical constraints of each hospital, in response to the bed capacity of each individual hospital. 

\subsection{Changing $\Rn$}

\begin{figure}[!htb]
    \centering
    \includegraphics[width=0.82\textwidth]{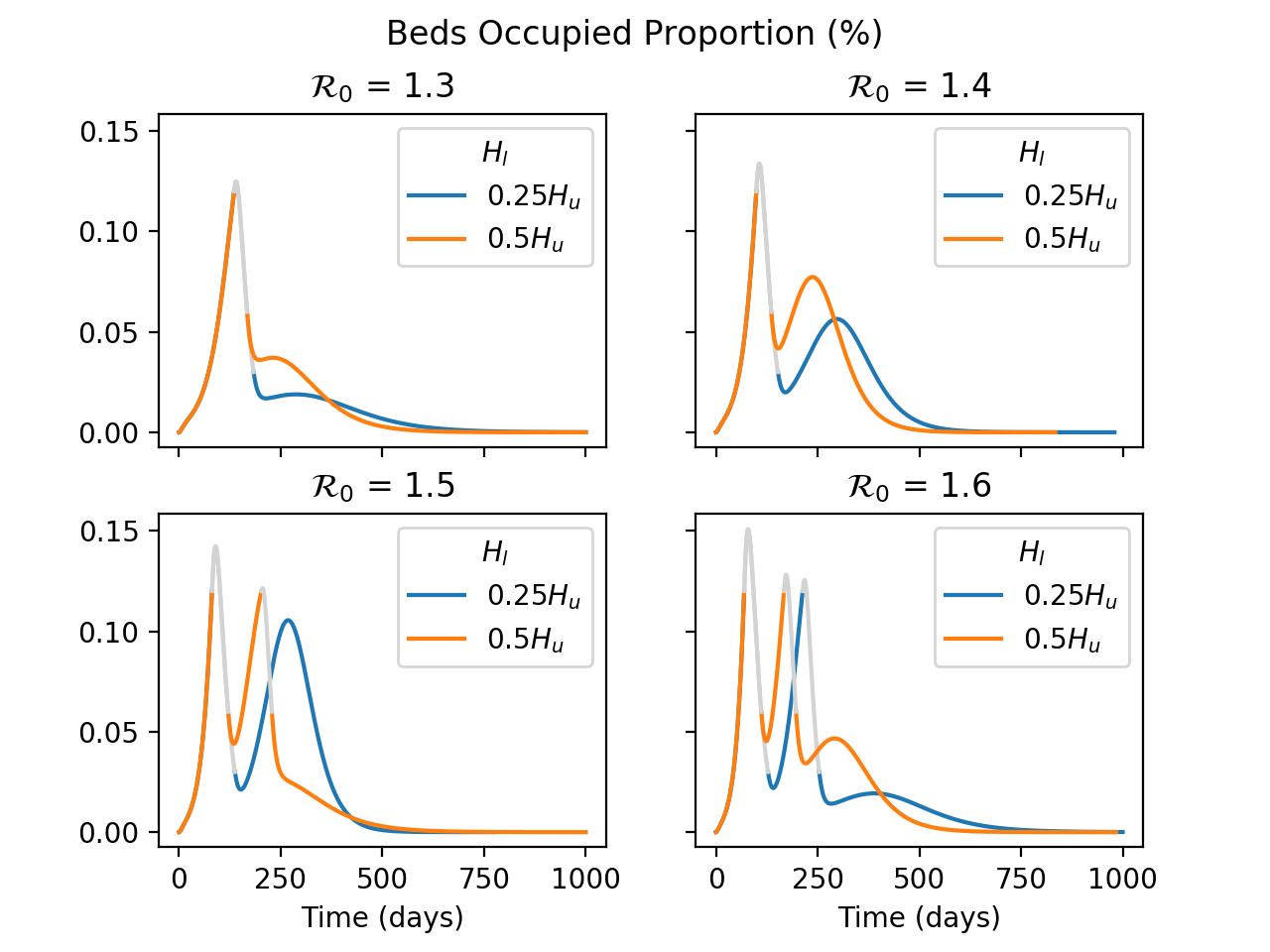}
    \caption{Percentage of patients in hospitals in the South East region using the hospital capacity intervention approach, changing $\Rn$ and fixing $H_l := 0.25 H_u$ or $H_l := 0.5 H_u$. The grey lines represent the times when the South East region is in an intervention. We note that we have truncated the simulation to make visualisation easier.}
    \label{fig:R0_intervention_H}
\end{figure}

\begin{figure}[!htb]
    \centering
    \includegraphics[width=0.82\textwidth]{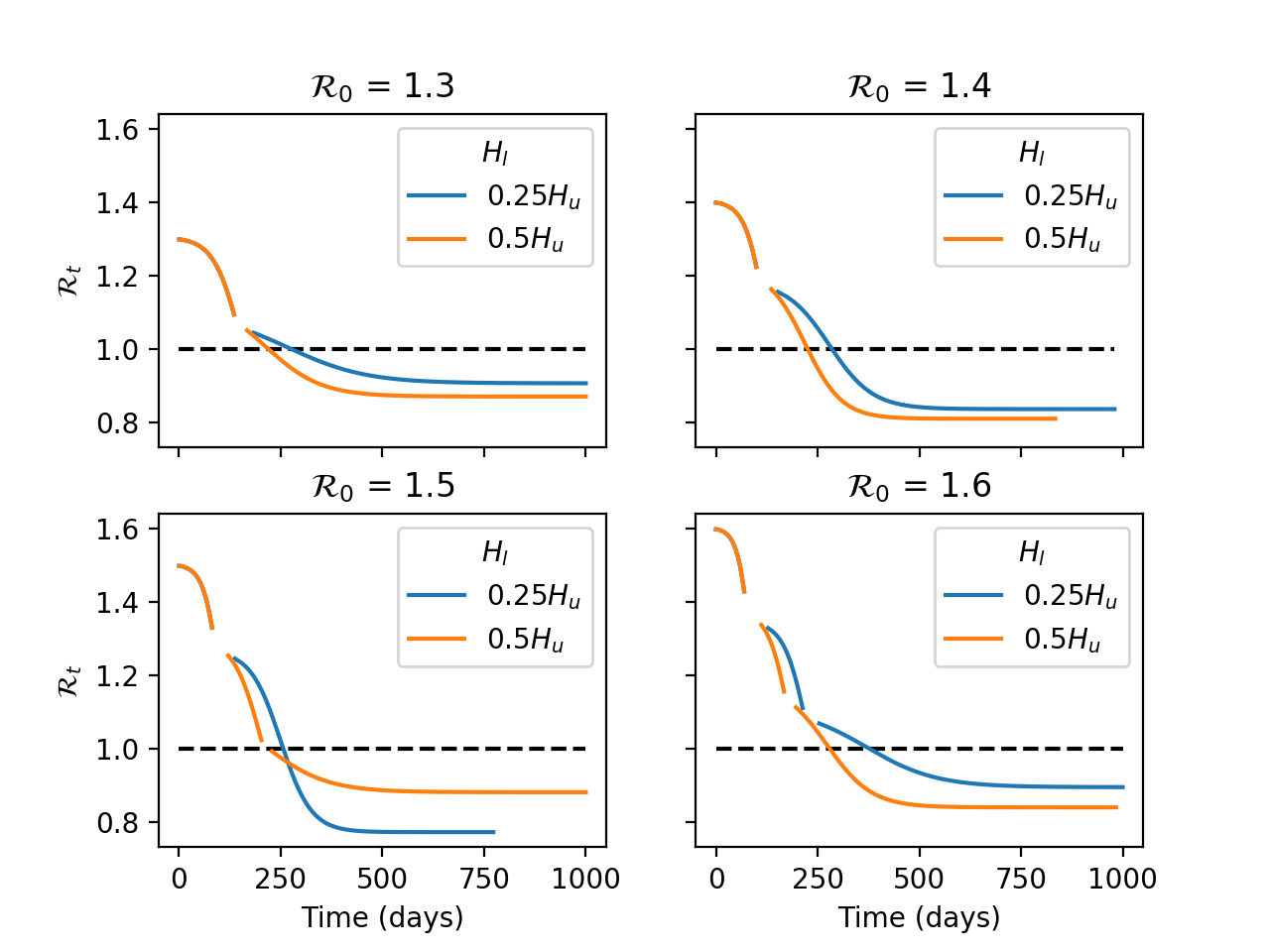}
    \caption{$\Rt$ in the South East region using the hospital capacity intervention approach, changing $\Rn$ and fixing $H_l := 0.25 H_u$ or $H_l := 0.5 H_u$. The gaps in the lines represent the times when the South East region is in an intervention. The black dashed line represents herd immunity $\Rt = 1$. We note that we have truncated the simulation to make visualisation easier.}
    \label{fig:R0_intervention_Rt}
\end{figure}

In this section, we will vary $\Rn$ whilst fixing $H_l := 0.25H_u$ or $H_l := 0.5H_u$. In Table \ref{tab:R0_intervention} we measure the percentage of dead individuals for each region for varying values of $\Rn$. In Figure \ref{fig:R0_intervention_H} we demonstrate the percentage of beds occupied in the South East region whilst in Figure \ref{fig:R0_intervention_Rt} we demonstrate the effective reproduction number over the simulation. The light grey colour in Figure \ref{fig:R0_intervention_H} represents the time when the regions are in an intervention, the same time period is represented by the gaps in the lines in Figure \ref{fig:R0_intervention_Rt}.

Comparing Table \ref{tab:dono} to Table \ref{tab:R0_intervention} we can see that the intervention has dramatically decreased the percentage of total deaths for larger values of $\Rn$, as expected. By looking at Figure \ref{fig:R0_intervention_H} we can see that as we increase $\Rn$, the number of interventions needed increases and also the length of the initial intervention increases. This is due to the number of future patients in the exposed compartment $E$ and the infectious compartment $I$. This is emphasised by the fact that the initial intervention is sooner for a larger value of $\Rn$ due to the increased average transmission. It can also be seen that the time between each intervention decreases as $\Rn$ increases. We can also see, by comparing Figure \ref{fig:donoRt} with Figure \ref{fig:R0_intervention_Rt}, that the epidemic actually lasts significantly longer with interventions included. These observations are realistically expected, however there are some results which are not necessarily expected or intuitive such as the percentage of dead individuals in the South East region decreasing from $\Rn = 1.5$ to $\Rn = 1.6$ when considering $H_l = 0.25 H_u$ but not decreasing for when $H_l = 0.5 H_u$. This is not specific to this value of $\Rn$, rather to the circumstance that this simulation finds itself after the final intervention. Namely, at this stage of the simulation for $\Rn = 1.6$, herd immunity has almost been reached, i.e. $\Rt$ is only slightly larger than 1. A value of $\Rt$ slightly larger than 1 means that although there is still an increase in the number of infectious individuals, the rate of that increase is much slower comparatively to an $\Rt$ value of, say, 1.5. At this stage, the number of incubating and infectious individuals in the case of $\Rn = 1.6$ are small which means that the final bump in the simulation for $\Rt = 1.5$ is much larger than the respective bump for $\Rt = 1.6$, as can be seen in Figure \ref{fig:R0_intervention_H}. One can see a similar behaviour happening between $\Rn = 1.4$ and $\Rn = 1.5$ when considering $H_l = 0.5H_u$. This interplay between parameter values, herd immunity and $\Rt$ is difficult to analyse and demonstrates that intuition is not necessarily enough when forecasting.

\begin{table}[!htb]
    \centering
    \caption{Percentage of dead individuals at the end of each simulation for each region. We fix $H_l := 0.25 H_u$ or $H_l := 0.5 H_u$ and vary $\Rn$.}
    \label{tab:R0_intervention}
    \begin{tabular}{lcccccc}
        \toprule
        & \multicolumn{2}{c}{England} & \multicolumn{2}{c}{South East} & \multicolumn{2}{c}{North West} \\
        \cmidrule{2-3}\cmidrule{4-5}\cmidrule{6-7}
        $\Rn$ & $\frac{H_l}{H_u} = 0.25$ & $\frac{H_l}{H_u} = 0.5$ & $\frac{H_l}{H_u} = 0.25$ & $\frac{H_l}{H_u} = 0.5$ & $\frac{H_l}{H_u} = 0.25$ & $\frac{H_l}{H_u} = 0.5$ \\
        \midrule
        1.3 & 0.797\% & 0.876\% & 0.735\% & 0.802\% & 0.843\% & 0.930\% \\ 
        1.4 & 1.048\% & 1.106\% & 0.978\% & 1.023\% & 1.073\% & 1.147\% \\ 
        1.5 & 1.263\% & 1.307\% & 1.176\% & 1.000\% & 1.291\% & 1.349\% \\ 
        1.6 & 1.139\% & 1.240\% & 1.069\% & 1.152\% & 1.473\% & 1.279\% \\ 
        1.7 & 1.311\% & 1.394\% & 1.234\% & 1.297\% & 1.315\% & 1.423\% \\ 
        1.8 & 1.460\% & 1.527\% & 1.369\% & 1.420\% & 1.467\% & 1.557\% \\ 
        1.9 & 1.584\% & 1.430\% & 1.481\% & 1.317\% & 1.599\% & 1.675\% \\ 
        2.0 & 1.689\% & 1.519\% & 1.323\% & 1.416\% & 1.712\% & 1.777\% \\
        \botrule
    \end{tabular}
\end{table}

\subsection{Changing $H_l$}

\begin{figure}[!htb]
    \centering
    \includegraphics[width=0.85\textwidth]{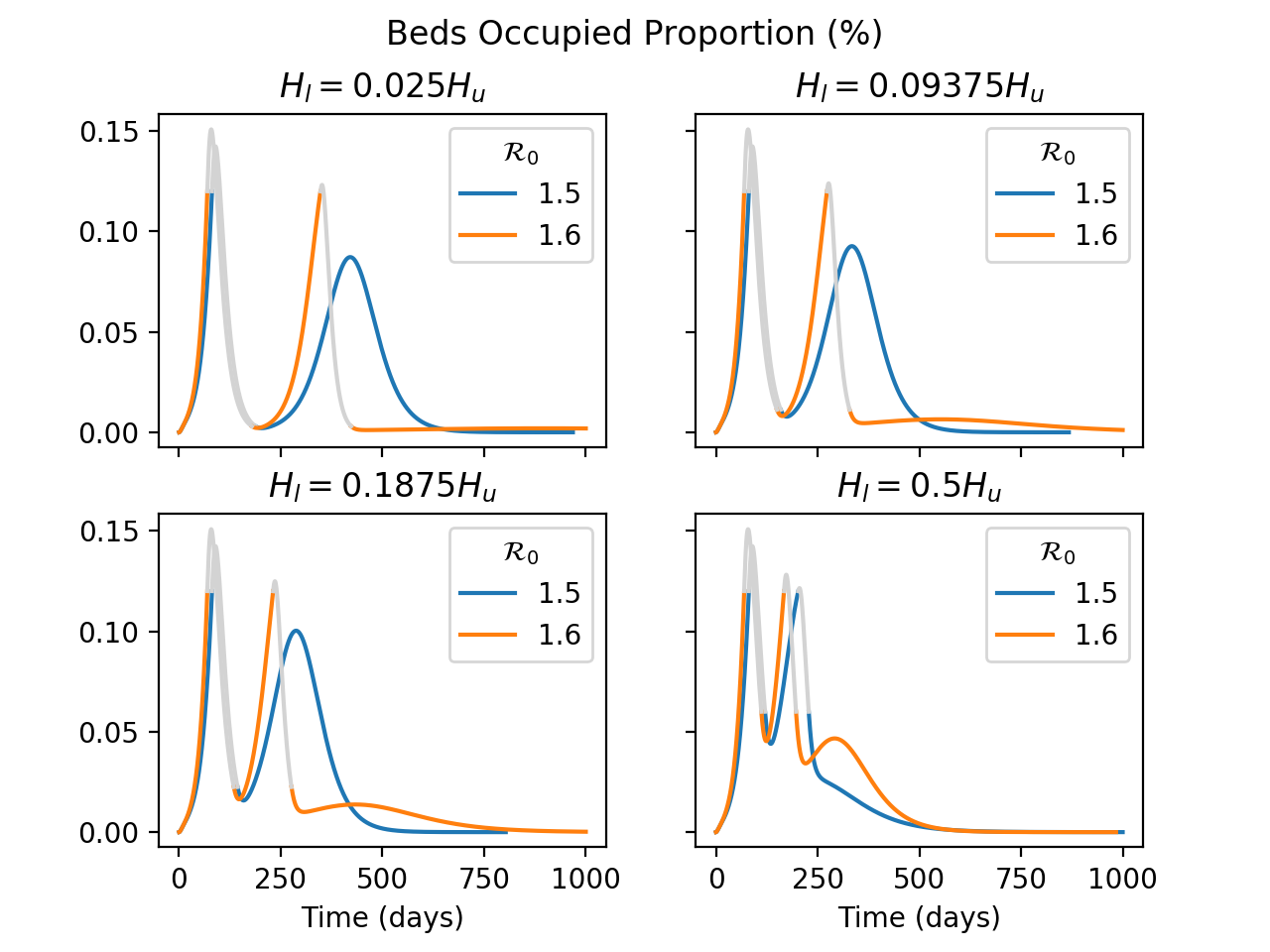}
    \caption{Percentage of patients in hospitals in the South East region using the hospital capacity intervention approach, changing $H_l$ and fixing $\Rn = 1.5$ or $\Rn := 1.6$. The grey lines represent the times when the South East region is in an intervention. We note that we have truncated the simulation to make visualisation easier.}
    \label{fig:Hl_intervention_H}
\end{figure}

\begin{figure}[!htb]
    \centering
    \includegraphics[width=0.85\textwidth]{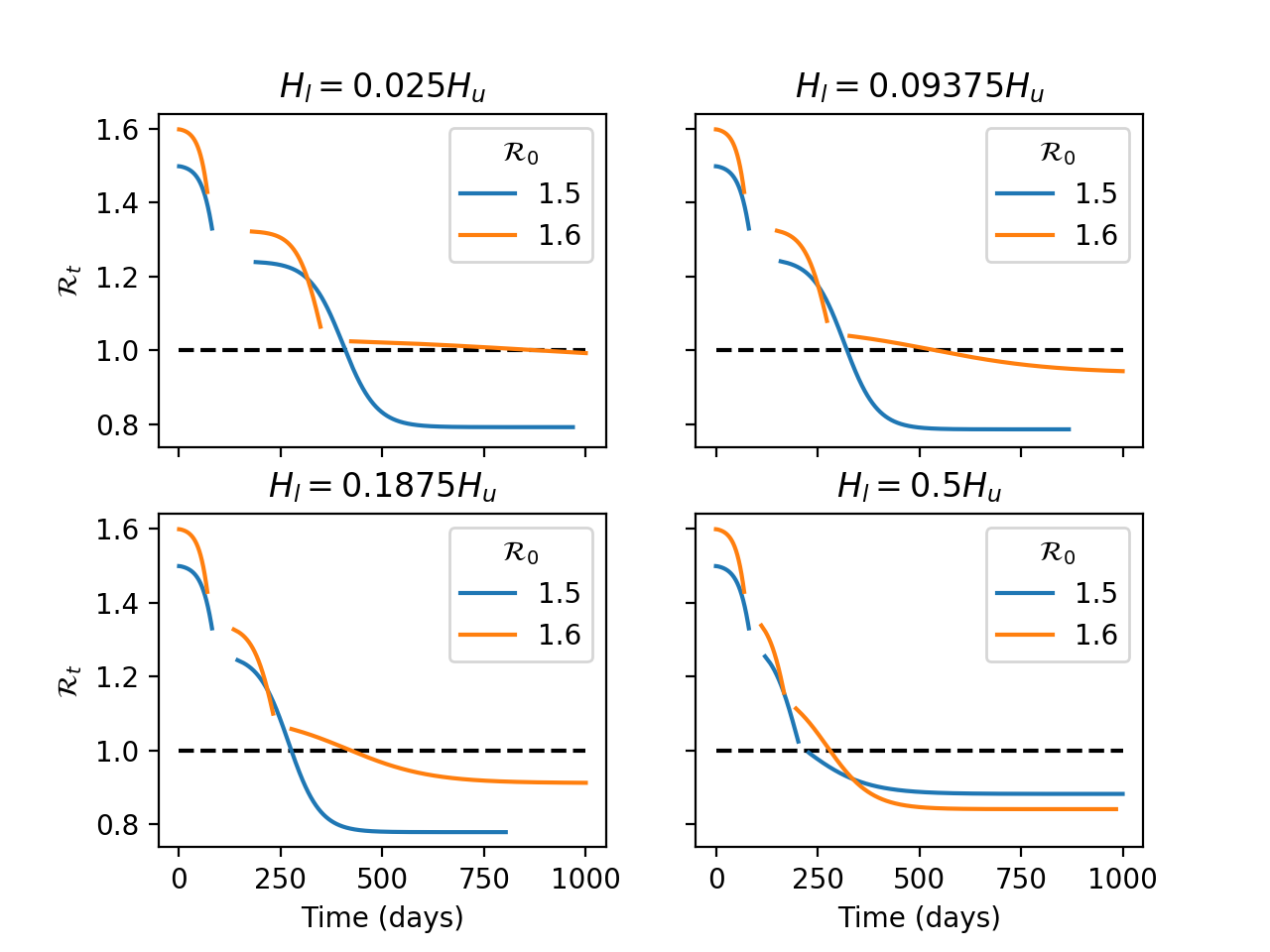}
    \caption{$\Rt$ in the South East region using the hospital capacity intervention approach, changing $H_l$ and fixing $\Rn = 1.5$ or $\Rn := 1.6$. The gaps in the lines represent the times when the South East region is in an intervention. The black dashed line represents herd immunity $\Rt = 1$. We note that we have truncated the simulation to make visualisation easier.}
    \label{fig:Hl_intervention_Rt}
\end{figure}

In this section, we will be experimenting by changing $H_l$, the lower threshold of patients that signals the ending of the intervention, and fixing $\Rn = 1.5$ or $\Rn = 1.6$. In Table \ref{tab:Hl_intervention} we measure the number of interventions needed, the length of each intervention (measured in days), the day of the initiation of each intervention and the percentage of the total deaths at the end of the epidemic. We demonstrate a few of the simulations in Figure \ref{fig:Hl_intervention_H} and we demonstrate the effective reproduction number over each outbreak in Figure \ref{fig:Hl_intervention_Rt}.

Comparing Table \ref{tab:dono} to Table \ref{tab:Hl_intervention} we can see that again the intervention has dramatically decreased the percentage of total deaths compared to the ``do-nothing" approach. Interestingly, changing the threshold $H_l$ mostly does not have that much of an effect on the percentage of total deaths, unlike the difference we saw in Table \ref{tab:R0_intervention} when changing $\Rn$, but it does have a large effect on the length of the outbreak. The effect of $\Rt$ is slightly different in this section, namely the sudden drop in the percentage of deaths is not caused primarily by a well-timed re-opening. Instead, the jump in percentage is due to the fact that an extra intervention has been used. Considering the South East region, and focusing on $\Rn = 1.5$, we can see that the percentage of deaths drops from 1.1682\% using $H_l = 0.1875H_u$ to 1.000\% using $H_l = 0.5H_u$, and by Figure \ref{fig:Hl_intervention_H} or \ref{fig:Hl_intervention_Rt} that there is one intervention using $H_l = 0.1875H_u$ whilst there is two interventions using $H_l = 0.5H_u$. 

The outcomes of our study so far imply that the timing and the lengths of interventions are extremely important. Getting closer to herd immunity when ending an intervention has the potential to save a huge number of lives. However, calculating $\Rt$ in real life is in general very challenging which leaves the process of timing for herd immunity difficult. One also notices that, although the percentage of total deaths decreases with an intervention, the length of most of the interventions is large due to the criteria set. This is mainly due to the fact that the average hospitalisation period is large and that the scenario we are simulating means that intervention will be in place until hospitals go from full capacity to between 2.5\% and 75\% capacity. Fortunately, we see that as the target capacity percentage increases, the percentage of total deaths does not increase dramatically, and the length of interventions decreases from the best part of 4 months to under 1 month. Similarly, as the outbreak progresses, one would expect the average hospitalisation length to decrease, since awareness of the disease and treatment gets better, as well as an increase in resources and the development of vaccines. This final point is important as it means realistic interventions can be implemented as circuit breakers and still maintain a large decrease in the number of total deaths. However, one aspect of this which is overlooked in this study is the potential for nosocomial outbreaks, whereby the probability of an outbreak increases with a larger number of infectious patients. 

\begin{table}[!ht]
    \centering
    \caption{Percentage of dead individuals at the end of each simulation for each region. We fix $\Rn = 1.5$ or $\Rn = 1.6$ and vary $H_l$.}
    \label{tab:Hl_intervention}
    \begin{tabular}{lcccccc}
        \toprule 
        & \multicolumn{2}{c}{England} & \multicolumn{2}{c}{South East} & \multicolumn{2}{c}{North West} \\
        \cmidrule{2-3}\cmidrule{4-5}\cmidrule{6-7}
        $\frac{H_l}{H_u}$ & $\Rn = 1.5$ & $\Rn = 1.6$ & $\Rn = 1.5$ & $\Rn = 1.6$ & $\Rn = 1.5$ & $\Rn = 1.6$ \\ 
        \midrule 
        0.025 & 1.221\% & 1.407\% & 1.145\% & 0.962\% & 1.232\% & 1.428\% \\ 
        0.0625 & 1.229\% & 1.096\% & 1.151\% & 0.987\% & 1.242\% & 1.435\% \\
        0.09375 & 1.235\% & 1.087\% & 1.155\% & 1.003\% & 1.251\% & 1.442\% \\ 
        0.125 & 1.240\% & 1.091\% & 1.159\% & 1.018\% & 1.259\% & 1.448\% \\
        0.1875 & 1.252\% & 1.113\% & 1.168\% & 1.0448\% & 1.275\% & 1.461\% \\
        0.25 & 1.263\% & 1.139\% & 1.176\% & 1.069\% & 1.291\% & 1.473\% \\ 
        0.5 & 1.307\% & 1.240\% & 1.000\% & 1.152\% & 1.349\% & 1.279\% \\ 
        0.75 & 1.180\% & 1.334\% & 1.082\% & 1.230\% & 1.242\% & 1.382\% \\
        \botrule
    \end{tabular}
\end{table}

\subsection{Capacity thresholds}
\label{minim_cap}

In this section we calculate what the highest capacity threshold $H_u$ is, for different values of $\Rn$, so that the hospitals do not go over their maximum capacity $H_{max}$. In particular, this approach can be used as an early warning system by outlining when interventions need to be enforced to maintain a manageable capacity. This approach can complement scenario-based forecasting approaches by giving a range of indicators of when to open up further capacity in hospitals or introduce an intervention which can be tracked against with incoming data daily. Ultimately, in practice, healthcare systems will want to utilise their capacity appropriately whilst not impacting the general public with an intervention, and so optimising the difference between the resource capacity and the maximum number of patients per parameter set is important.

\begin{figure}[!htb]
    \centering
    \includegraphics[width=0.6\textwidth]{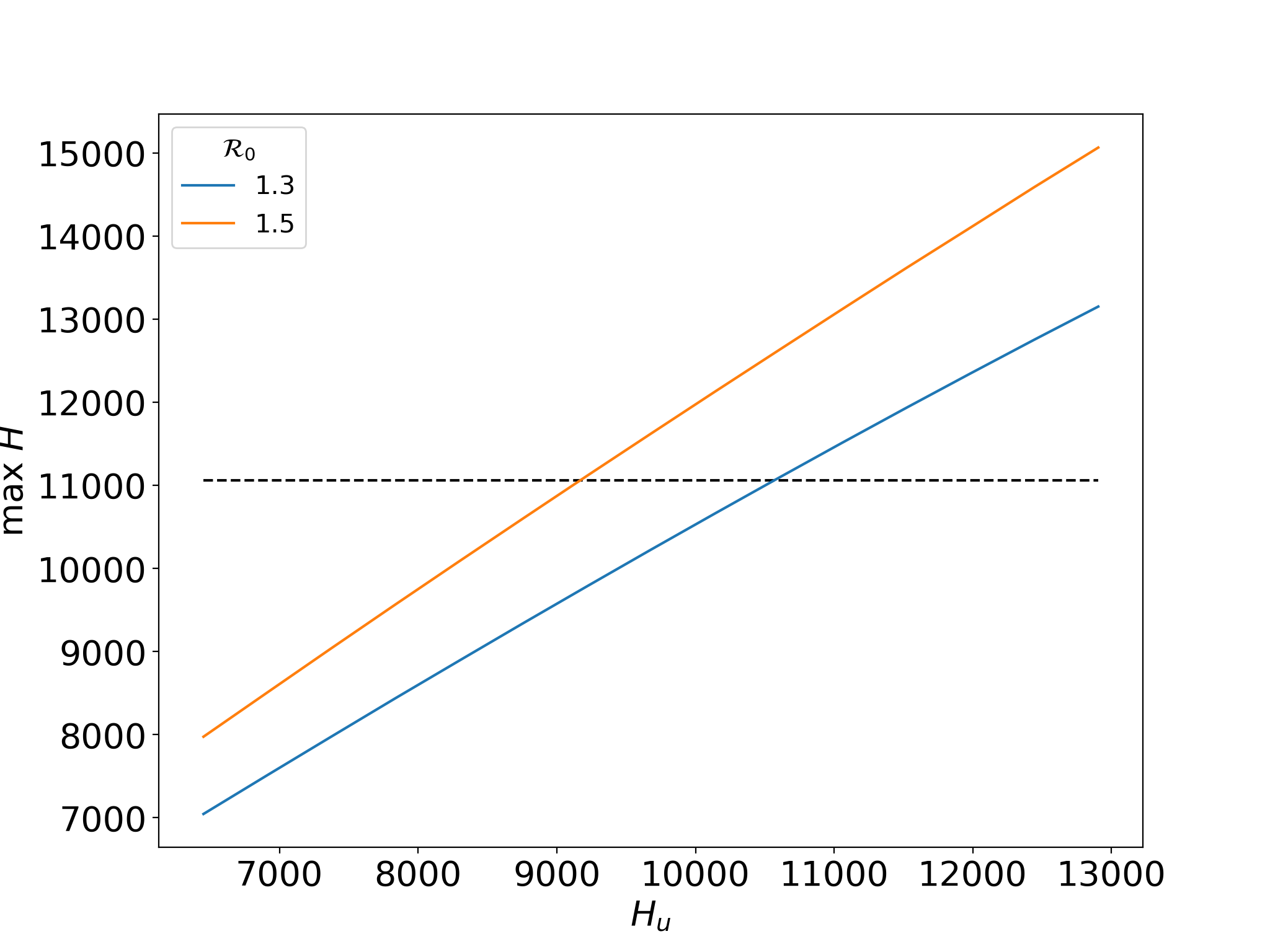}
    \caption{Comparison of $H_u$ against the $\max H$ for different values of $\Rn$ in the South East region. The black dashed line represents $H_{max}$.}
    \label{fig:capacity_func}
\end{figure}

In this section, we only consider the situation of finding $H_u$ using one intervention, that is to say \eqref{ell} becomes
\begin{align*}
    \ell := [\ell = 0][H(t) > H_u],
\end{align*}
with \eqref{bbar} the same. We only need to consider one intervention here because in most situations, once the hospitals have opened back up again, there will be another spike in capacity, however one can reason that this second spike is essentially the same as the first spike, just with slightly different parameters. For a specified value of $\Rn$, we look to find the root of the following function
\begin{align*}
    \mathcal{L}(H_u;H,H_{max}) := \max_{t>0} H(t) - H_{max},
\end{align*}
where we note that $H$ depends on $H_u$. Since $H$ is continuous, this function is continuous and will always have a root provided $\Rn$ is chosen high enough such that the maximum of $H$ without an intervention is at least as large as $H_{max}$. Figure \ref{fig:capacity_func} depicts the situation where $\Rn$ is chosen appropriately for the South East region. For this section, we take $H_{max} := 0.0012 N$. 

\begin{figure}[!htb]
    \centering
    \includegraphics[width=0.6\textwidth]{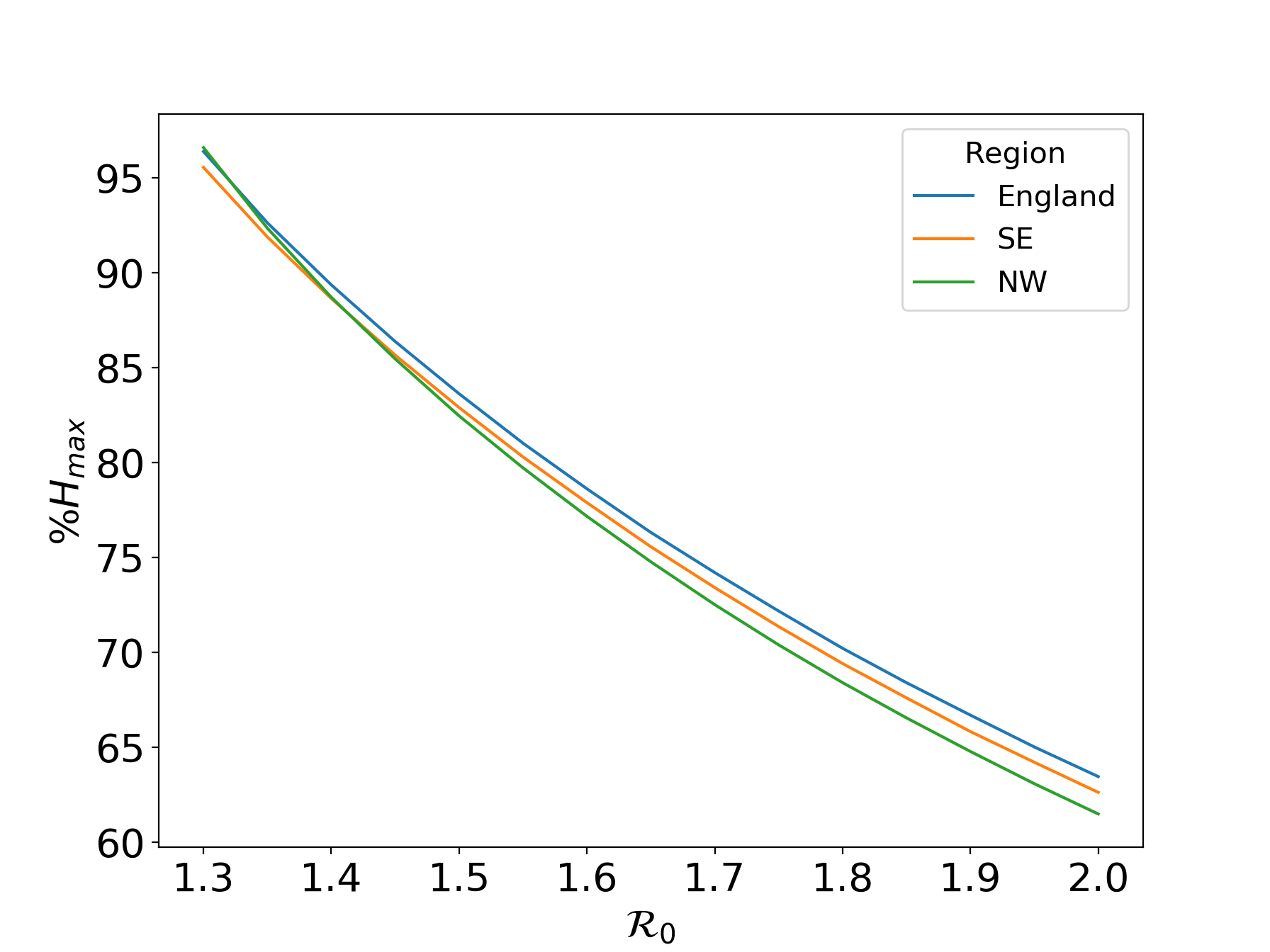}
    \caption{The percentage of $H_{max}$ corresponding to the roots of $\mathcal{L}$ for different values of $\mathcal{R}_0$, where the roots correspond to values of $H_u$. Here North West is abbreviated to NW and South East is abbreviated to SE.}
    \label{fig:opt_capacity}
\end{figure}

We demonstrate the results in Figure \ref{fig:opt_capacity}. This figure should be read in the following way: if we know $\Rn$, then we look to determine that an intervention should be initiated when the healthcare system reaches a certain percentage of the maximum resource capacity. We can see that as $\Rn$ increases, the cap on the occupancy for the intervention decreases, which is to be expected. This measurement is useful, as, depending on the value of $\Rn$, one can calculate what the cap is on occupancy and thus can judge when to take steps to initiate an intervention. 

\section{Agent-based Approach}
\label{sec:ABM} 

In this section we look to explore the use of an agent-based model for hospital occupancy. Similar to standard compartmental models, agent-based models (also known as individual-based models) are versatile in what they can model, being used in areas such as biology, engineering, politics and economics \cite{DD19,PSR98,RG19,SJNSS18}. The agent-based approach considers agents and their behaviours, whilst the standard compartmental approach, also referred to as equation-based modelling in the literature, focuses on observables and equations. There are direct parallels between the two approaches, as we will demonstrate, but the manifestation of the approaches can lead to different results. Due to the nature of infectious diseases, having recognisable transmissions between distinct states, agent-based approaches have slowly been making a surge to becoming the standard method for mathematical epidemiological modelling \cite{D05,LS12,PD09}. The agent-based approach can characterise variable contacts between agents in a way which is difficult to describe using the equation-based approach, since the agent-based approach is more flexible in its description due to the use of probability. The famed Imperial College Model \cite{FLNal20}, the implementation is known as CovidSim, is an agent-based model, based on previous works conducted on modelling influenza pandemics \cite{FCCFRMIB05,FCFCCB06,HFELCLXFVAGal08}. Other widely used agent-based models, particularly for the modelling of COVID-19, are covasim and OpenABM-Covid19 \cite{AHWLPWESRD21,HPNKWHLBZS21,KSMARHNRCSH21,PKSMKVB20}. We look to compare the outcomes of the previous sections, using the same parameters, in the agent-based approach and demonstrate that the agent-based approach offers more flexibility in devising interventions. 

\subsection{Model formulation and validation}

\begin{figure}[!hbt]
    \centering
    \includegraphics[width=\linewidth]{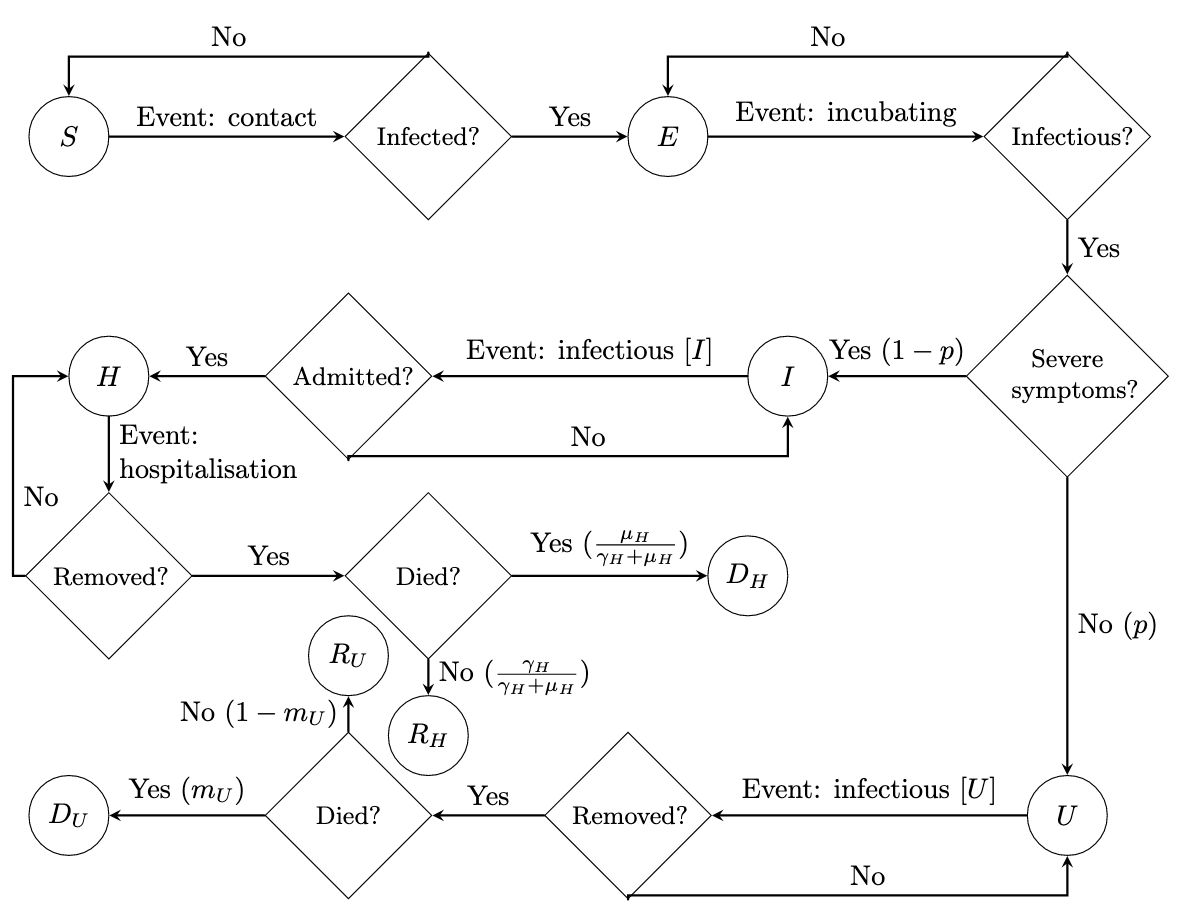}
    \caption{Schematic representation of the flowchart which results in the agent-based mathematical model.}
    \label{fig:abm_model}
\end{figure}

In this section we will present the agent-based version of the equations in \eqref{Seq}--\eqref{DHeq}. In particular, we look to draw parallels between the two schematics in Figures \ref{fig:model} and \ref{fig:abm_model}. In a similar manner to Figure \ref{fig:model}, Figure \ref{fig:abm_model} describes the way an agent moves between states and should be read as a flowchart, where we have labelled the states similarly to demonstrate some of the parallels. In Figure \ref{fig:abm_model} the circles represent the state an individual can be, and the diamonds represent a decision. The arrows from a states to a decisions are an event in the epidemic, and the arrows from a decision to a state or decision are the outcome of the decision. Some arrows have an outcome of a decision with a parameter in parentheses next to them, these parameters are probabilities of traversing a different section of the flowchart and represent the same parameter as when there is a fork in states in Figure \ref{fig:model}, such as the probability of not going to hospital $p$. Some arrows have an event with a state in brackets next to them, these are to distinguish between the different rate parameters as the states represent similar characteristics, such as the multiple types of being infectious $U$ and $I$. 

For ease of exposition, we consider the time unit of the simulation to be days. Each event decision is decided by drawing one pseudo-random uniform number from the interval $[0,1]$, let us call this value $x$, and comparing it to the probability of the event on a given day, let us call this value $d$. We can describe this process as the following categorical function
\begin{align*}
    f(x;d) := [x \leq d]\text{``Yes"} + [x > d]\text{``No"}. 
\end{align*}
We note that for each event, a new $x$ is generated, and each $d$ is different. The events can be split into two categories, transmission and progression. We begin by describing the transmission event, event: contact. For ease of exposition, each day we assume that all agents make a fixed $C$ random contacts with a probability of a successful transmission $a$, where $\beta = Ca$. For a susceptible agent, of those $C$ contacts, we denote $\kappa$ to be the number of contacts they make with an infectious agent, namely an agent in state $U$ or $I$. Then, we have that
\begin{align*}
    d = \prob(X \leq \kappa) = 1 - (1-a)^\kappa,
\end{align*}
where $X \sim Geo(a)$ and $X$ has support on $\{1,2,\dots\}$. The remaining events are all progression events, whereby the length of time an agent has spent in a state is important. We will describe ``event: incubating", but note that the same description can be used for each of the events using their respective rate parameters. Let $l$ represent the number of days an agent has spent in the $E$ state, then we have
\begin{align*}
    d = \prob(Y_E \leq l) = 1 - e^{-\gamma_E l},
\end{align*}
where $Y_E \sim Exp(\gamma_E)$. One can see that event: infectious $[U]$ gives $Y_U \sim Exp(\gamma_U)$, event: infectious $[I]$ gives $Y_I \sim Exp(\gamma_I)$, and event: hospitalisation gives $Y_H \sim Exp(\gamma_H + \mu_H)$. We note that this is not the standard methodology for agent-based models, typically the length of stay for each agent is not stochastically generated, but instead each day an agent runs a Bernoulli trial using the probability associated to the event of progression. In our framework, it is obvious to see that compartments are exponentially distributed, and there is an obvious way to change that assumption if needed without having to describe the underlying mathematics differently. For example, if it was known that the incubation event was in fact akin to an Erlang distribution rather than an Exponential, then we can update the agent-based approach quite simply by updating the random variable $Y_E$, however in the equation-based approach we would need to add more compartments (the additional number depending on the distribution's shape parameter) \cite{CDE18}. 

\begin{figure}[!htb]
    \centering
    \includegraphics[width=0.6\textwidth]{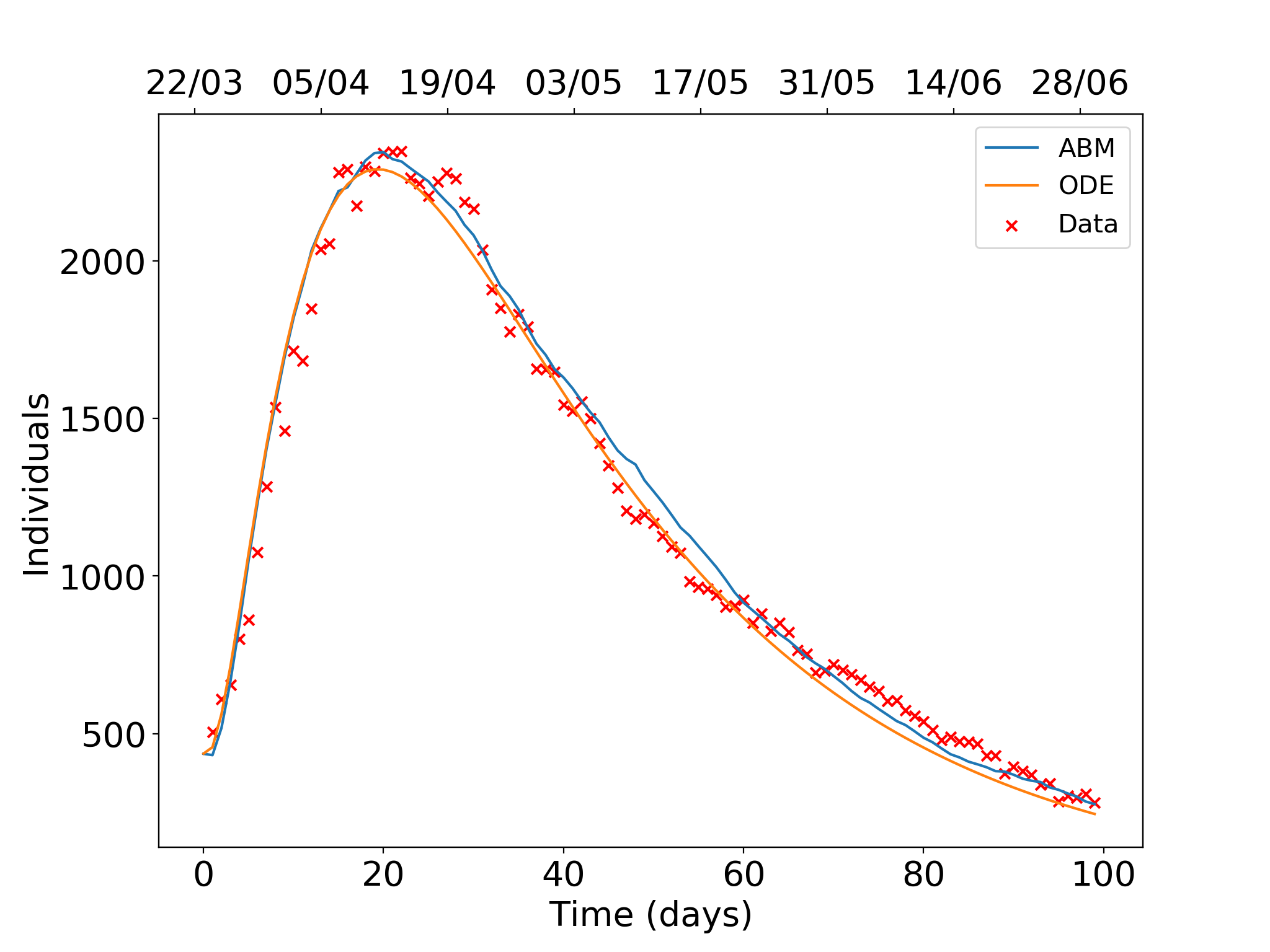}
    \caption{Comparison of the agent-based approach to the standard deterministic SEIR-D approach using the South East parameters. Both approaches match the data closely throughout the time period.}
    \label{fig:abm_beds_comp}
\end{figure}

For the remainder of this study, we will focus only on the outcomes using the South East parameters. In order to compare the agent-based approach to the standard deterministic SEIR-D approach, we run 100 Monte Carlo simulations and take the average number of agents in each state for each day. We note that typically 100 Monte Carlo simulations are not enough but we see very little variation in the average from using as low as 5 iterations. Due to the stochastic and numerical discretisation, we adjust the rate parameters by adding a correction term in the form of
\begin{align}
    c(\gamma; \dt) = \frac{\gamma^2 \dt}{2 - \gamma \dt}, \label{correction_term}
\end{align}
where $\dt$ denotes the time interval being considered in ratio of days, e.g. $\dt=0.5$ would be considering half days, see \ref{justification} for justification. We note that $\dt$ needs to be chosen appropriately so that the denominator in $c$ is non-zero, however this is a reasonable assumption since typically $\gamma < 1$ (as it is a rate and its reciprocal in normally larger than one) and $\dt \leq 1$ since we typically consider simulating each day or several simulations per day. Using the parameters defined in Table \ref{tab:all_params}, with the associated fitted initial conditions, taking $\dt = 0.25$ results in Figure \ref{fig:abm_beds_comp}, whereby we see a clear agreement between the two approaches and the data. We note that this result is also due to the fact that $N$ is large, and the initial conditions are suitably large, which combined are often called the mean-field assumption (or the thermodynamic limit). From here onwards, we will fix $\dt = 1$ for speed but we note in \ref{justification} that using $\dt = 1$ and the correction term results in outputs sufficiently close to the SEIR-D model. 

\subsection{Intervention scenarios}

In this section, we look to apply the techniques and parameter changes in Section \ref{sec:EBM} to the agent-based approach. We know that $\beta$ is multiplicatively formed of the average probability of transmission $a$, and the average number of contacts per day $C$. When changing the transmission rate, $\beta$, we will consider two different approaches. We look to investigate whether just changing the probability of transmission $a$ or average number of contacts per day $C$ have different impacts on the outcome of the epidemic. One can reason that a change in probability could be due to new strains of an infectious disease or a change in public behaviour, such as wearing masks or increased personal hygiene. Changing the number of contacts could be due to a national level intervention, such as lockdowns or closing schools. We note that we whilst it would be scientifically ideal to keep $a$ fixed, due to the integer nature of contacts $a$ has to be adjusted to maintain the same $\Rt$ value across the measurements, thus making sure the tables are comparable.

By using the ``do-nothing" approach, we see that in fact changing either parameter does not make an overall difference to the outputs, which is to be expected given the mean-field limit of the agent-based approach to the compartmental model. From here onwards, we will fix $C = 9$ contacts per day and only adjust $a$. The results in Tables \ref{tab:donoABM_a} and \ref{tab:donoABM_c} also match up to the results in Table \ref{tab:dono}. Applying the agent-based approach to the intervention scenarios described in Section \ref{intervention}, we see that the results in Table \ref{tab:R0_intervention_abm} match with the results in Table \ref{tab:R0_intervention} as expected, even maintaining the interesting behaviour around the timing of herd-immunity. 

\begin{sidewaystable}
\sidewaystablefn%
\begin{center}
\begin{minipage}{\textheight}
\caption{Measurements of the ``do-nothing" approach using the agent-based approach with fixed $C$.}
    \label{tab:donoABM_a}
\begin{tabular*}{\textheight}{@{\extracolsep{\fill}}cccccc@{\extracolsep{\fill}}}
\toprule%
$\Rn$ & $C$ & $a$ & Max beds occupied (\%$N$) & Peak of beds occupied (day) & Dead individuals (\%$N$) \\
\midrule
1.3 & 9 & 0.0300 & 0.158\% & 172 & 1.044\% \\  
        1.4 & 9 & 0.0326 & 0.241\% & 149 & 1.253\% \\  
        1.5 & 9 & 0.0349 & 0.328\% & 130 & 1.427\% \\  
        1.6 & 9 & 0.0373 & 0.415\% & 117 & 1.569\% \\ 
        1.7 & 9 & 0.0396 & 0.502\% & 108 & 1.688\% \\  
        1.8 & 9 & 0.0419 & 0.585\% & 100 & 1.784\% \\ 
        1.9 & 9 & 0.0443 & 0.661\% & 93 & 1.869\% \\ 
        2.0 & 9 & 0.0466 & 0.739\% & 89 & 1.938\% \\
\botrule
\end{tabular*}
\end{minipage}
\end{center}
\end{sidewaystable}

\begin{sidewaystable}
\sidewaystablefn%
\begin{center}
\begin{minipage}{\textheight}
\caption{Measurements of the ``do-nothing" approach using the agent-based approach with varied $C$.}
    \label{tab:donoABM_c}
\begin{tabular*}{\textheight}{@{\extracolsep{\fill}}cccccc@{\extracolsep{\fill}}}
\toprule%
$\Rn$ & $C$ & $a$ & Max beds occupied (\%$N$) & Peak of beds occupied (day) & Dead individuals (\%$N$) \\
\midrule
1.3 & 9 & 0.0300 & 0.158\% & 172 & 1.045\% \\ 
        1.4 & 10 & 0.0293 & 0.241\% & 147 & 1.256\% \\ 
        1.5 & 11 & 0.0286 & 0.328\% & 131 & 1.422\% \\ 
        1.6 & 12 & 0.0279 & 0.415\% & 117 & 1.568\% \\ 
        1.7 & 13 & 0.0274 & 0.501\% & 108 & 1.689\% \\ 
        1.8 & 14 & 0.0269 & 0.585\% & 100 & 1.788\% \\ 
        1.9 & 15 & 0.0266 & 0.664\% & 94 & 1.870\% \\  
        2.0 & 16 & 0.0262 & 0.739\% & 88 & 1.940\% \\
\botrule
\end{tabular*}
\end{minipage}
\end{center}
\end{sidewaystable}

\begin{table}[!htb]
    \centering
    \caption{Percentage of dead individuals at the end of each simulation for the South East region using the agent-based approach. We fix $H_l := 0.25 H_u$ or $H_l := 0.5 H_u$ and vary $\Rn$.}
    \label{tab:R0_intervention_abm}
    \begin{tabular}{ccc}
        \toprule
        $\Rn$ & $\frac{H_l}{H_u} = 0.25$ & $\frac{H_l}{H_u} = 0.5$ \\
        \midrule 
        1.3 & 0.742\% & 0.802\% \\ 
        1.4 & 0.992\% & 1.023\% \\ 
        1.5 & 1.185\% & 1.003\% \\ 
        1.6 & 1.067\% & 1.138\% \\ 
        1.7 & 1.230\% & 1.288\% \\ 
        1.8 & 1.356\% & 1.414\% \\ 
        1.9 & 1.476\% & 1.310\% \\ 
        2.0 & 1.309\% & 1.386\% \\
        \botrule
    \end{tabular}
\end{table}

\subsection{Capacity thresholds}

Now, using the agent-based approach, we look to explore and compare against the results presented in Section \ref{minim_cap}. It must be noted that in the deterministic setup we calculated the threshold of the maximum number of patients in hospital before an intervention is needed to guarantee that capacity is not breached. In this section, we want to calculate the number of Monte Carlo realisations that still result in a breached capacity using the calculated deterministic threshold and how the number of realisations that breach capacity decreases when that threshold is reduced. In particular, we look to do a more realistic investigation to the thresholds using the in-built stochasticity of the agent-based approach. This investigation can act as some sort of buffer for hospital management to understand how close to the threshold they can get before having to make changes. We note here that the agent-based approach tends to overestimate its deterministic counterpart when $\dt = 1$, as approaches the deterministic solution from above when reducing $\dt$, as seen in Figure \ref{fig:correction_conv} in \ref{justification}. This implies that the results will be skewed towards a breach in the threshold. In this investigation we measure three metrics, the number of Monte Carlo realisations that go over $H_{max}$, the average maximum amount the realisations go above $H_{max}$ (to give an idea of how severely the threshold is breached), and the average maximum difference of the simulated hospital capacity $H$ from $H_{max}$ across all realisations, given a value of $H_u$ associated to a value of $\Rn$ from Figure \ref{fig:opt_capacity}. We measure these metrics against the percentage of $H_u$ which initiates an intervention, ranging from 90\% of $H_u$ to 100\% of $H_u$. We make several observations from the results in Table \ref{tab:capacity_abm}: 
\begin{enumerate}
    \item The higher the value of $\Rn$, the higher the chance the system will be breached;
    \item The higher the value of $\Rn$, the lower the percentage of the threshold needs to be considered to reduce the chance of a breach; 
    \item The higher the value of $\Rn$, the larger the system breach will be on average. 
\end{enumerate} 
Intuitively, these observations make sense and should not necessarily come as a surprise. However, what is useful about this study is understanding what percentage of $H_u$ is needed to reduce the breaches. This result and process can be used by hospital administrations, planners and managers to prepare for an incoming wave and set aside surge capacity in case of a breaching realisation. 

\begin{sidewaystable}
\sidewaystablefn%
\begin{center}
\begin{minipage}{\textheight}
\caption{Understanding the breaching of the maximum capacity using the agent-based approach by changing the threshold level determined in Section \ref{minim_cap}.}
    \label{tab:capacity_abm}
\begin{tabular*}{\textheight}{@{\extracolsep{\fill}}ccccccc@{\extracolsep{\fill}}}
\toprule
& \multicolumn{2}{c}{\makecell{Number of breaching \\ realisations}} & \multicolumn{2}{c}{\makecell{Average maximum amount \\ over $H_{max}$ (\%)}} & \multicolumn{2}{c}{\makecell{Average maximum difference \\ from $H_{max}$ (\%)}} \\
         \cmidrule{2-3}\cmidrule{4-5}\cmidrule{6-7}
        \%$H_u$ & $\Rn = 1.3$ & $\Rn = 2$ & $\Rn = 1.3$ & $\Rn = 2$ & $\Rn = 1.3$ & $\Rn = 2$ \\ 
\midrule
90\% & 0 & 20 & 0 & 1.934\% & -6.421\% & -2.900\% \\ 
        95\% & 0 & 80 & 0 & 2.780\% & -2.060\% & 1.017\% \\ 
        97.5\% & 32 & 90 & 0.863\% & 4.437\% & -0.108\% & 3.793\%\\
        100\% & 90 & 96 & 2.316\% & 6.625\% & 1.019\% & 5.960\% \\ 
\botrule
\end{tabular*}
\end{minipage}
\end{center}
\end{sidewaystable}

\section{Limitations of the modelling approach and their mitigation}
\label{sec:further}

In this study, we assumed that the average probability of going to hospital is the same throughout the different regions due to problems with parameter estimation and the initial conditions. We speculate that the issue of estimating the initial conditions without keeping $p$ fixed is solvable by reformulating the non-linear initial value problem into a non-linear boundary value problem, where the data is used directly in the model rather than an attribute of the fitting process. We have so far demonstrated how the boundary value problem would be conceived for a simple SIR model and prove existence and uniqueness of the problem for fixed parameters, in the future we look at parameter identifiability, parameter estimation and efficient numerical algorithms using this method \cite{CWVDM21}. Methods such as the shooting method, nonlinear least squares of the equivalent initial value problem, or numerical continuation are the standard methods of choice \cite{AG12,B81,HMTMW00,RHCC07}, however it is not clear which solution method is the most appropriate for the solution to the nonlinear boundary value problems that can be derived from models used in epidemiology. 

The current research into vaccines will prove pivotal in the role to reduce COVID-19 resurgences in the future \cite{BI20,BRKLCGL21,LSHH20,MHTDK21,SWBal20}. Mathematically, the addition of a vaccine into the model has been undertaken in previous works of similar nature \cite{GTNT07,MHTDK21}, but as with the other aspects we have mentioned, getting reliable data and understanding the appropriate mathematical assumptions remains the challenge. The ideal situation is that the model captures the reality of the vaccination strategy, namely vaccination rates and parameters are based on live information and data. Whilst the number of COVID-19 vaccinations is published at least on a daily basis, the difficulty lies in understanding the vaccination status of those in hospital. Vaccination assumptions and contributions to models often take one of two forms: the sterilizing vaccine and the leaky vaccine approach. The sterilizing approach assumes that a proportion of those receiving vaccines gain (some form of) immunity whilst the others do not. This style of approach lends itself well to agent-based approaches due to the application of a criteria to specific individuals (agents). The leaky vaccine approach considers that all individuals who receive a vaccine receive a proportion of protection, this style of approach lends itself well to equation-based models as it considers the population as an aggregated state. 

An oversight of the work here is that the county is not homogeneous with respect to age. Different age-groups have different social structures, responsibilities - such as working, school life, or family, and different responses to a COVID-19 infection. In this work, we have not explored the impact of inhomogeneity within the population on hospital demand and capacity, in principle due to the lack of publicly available data. In general, the intervention we impose on our system is a total lockdown of all ages, similar to the national lockdown during the first wave, however utilising age-groups within the model will allow for dedicated forecasting into the effect some social events like schools opening or returning to offices will have on interventions \cite{DBDal20,KKCal20,NMDal20}. 

Sticking to the models presented here, we can take steps forward to consider that maximising capacity and having longer lockdowns might not be more beneficial than small ``circuit breaker" lockdowns when one considers the cost of hospital use and the local economy. For example, by associating a cost to hospital usage or to a lockdown in general, we can find the maximum capacity threshold to go into a lockdown such that, for a specific value of $\Rn$, we minimise the total costs by using some of the measurements we presented here such as the length of lockdown, number of lockdowns, and the time until the peak from the initiation of a lockdown. A similar study was conducted in \cite{EHMal20} where they calculated the cost of capital (e.g. extra hospitals, provision of hand-washing stations) and one-time costs (e.g. hiring consultants to adapt policy, prepare online training courses), the cost of commodities (e.g. extra single use masks, specific increase in drugs) and the cost of human resources (e.g. extra doctors, extra cleaners) and combined it with the the estimated number of cases from the Imperial College model \cite{FLNal20} after four weeks and twelve weeks, using an increase and decrease of 50\% transmission rate for an interval of costs. Since their model is on a national scale and uses national derived parameters, one can extend their modelling approach to regional and local levels by using our fitted model. 

From a practical perspective, the next question to ask is: now we know what levels the hospital can take, what about the recovery procedure? It is well known that recovering from COVID-19 is not as easy as recovering from, say, the common cold \cite{CBLal20,YWVal20}, some people completely recover but invasive treatment may have caused further complications, whilst some people may continue to show effects of COVID-19, namely suffering from long COVID \cite{Al20,Sh20,SMVGPBPCKAC21}. In this sense, it is natural to extend the model here to include, what the NHS Clinical Commissioning Group label as, the discharge pathways, which describe the nature of the discharge of a patient and what recovery services they will need. Each pathway describes the level of need of a discharged patient, each level having an associated requirement and cost. Hence the following question arises: what will the burden to healthcare across the country be in one year, five years, and so on? Understanding the pressure on discharge pathways due to COVID-19 may give an indication on recovery costs post COVID-19 infection and/or hospitalisation. 

\section{Conclusion}
\label{sec:conc} 
In this study, we have presented a computational approach for measuring the impact of healthcare demand and capacity due to surges in COVID-19 infections and hospitalisations. We have used the notion of hospital capacity as a measure for exploring intervention scenarios that will allow hospitals to predict and forecast when demand and capacity are close to being breached and therefore allow resource allocations where necessary. 

The key findings are:
\begin{itemize}
    \item we have demonstrated that interventions will make a significant impact on the percentage of individuals who will die as a result of COVID-19;
    \item we have described an easily definable and understandable method of introducing an intervention which does not depend on prior knowledge of when the peak of infections will be;
    \item we have described parallels between equation-based modelling and agent-based modelling and demonstrated that the interventions work in both frameworks; 
    \item we have demonstrated that only changing the number of contacts and only changing the probability of transmission provide the same outputs;
    \item we have shown that, although a threshold to hospital capacity to deny a breach is calculated in the deterministic setup, that threshold needs to be decreased in order to account for stochastic perturbation of a realisation of an epidemic to not go over capacity. 
\end{itemize}

Our approaches are built around using a simple SEIR-D model coupled with novel statistical methods for parameter estimation and reframing the system, with the found parameters, in the agent-based approach to allow us to explore various plausible hypothetical scenarios that are of interest to the NHS local healthcare management teams and death management teams in local authorities \cite{CVAal20}. The theoretical and computational approach has a strong interplay between data and the model, whereby data drives the optimal parameter estimates and these in turn drive model predictions through dynamic models. 

\textbf{Acknowledgements and software:} The authors would like to acknowledge the use of Python, in particular the NumPy, Matplotlib, pandas and SciPy libraries \cite{numpy,matplotlib,pandas,scipy}. This work was partly supported by Brighton and Hove City Council, East and West Sussex County Councils and the NHS Sussex commissioners (JVY), and partly supported by the Global Challenges Research Fund through the Engineering and Physical Sciences Research Council grant number EP/T00410X/1: UK-Africa Postgraduate Advanced Study Institute in Mathematical Sciences. AMa acknowledges partial support from  the Health Foundation (1902431), the NIHR (NIHR133761) and by an individual grant from the Dr Perry James (Jim) Browne Research Centre on Mathematics and its Applications (University of Sussex). AMa is a Royal Society Wolfson Research Merit Award Holder funded generously by the Wolfson Foundation (2016-2021). AMa is a Distinguished Visiting Scholar to the Department of Mathematics, University of Johannesburg, South Africa. All authors acknowledge the continued support and collaboration of Brighton and Hove City Council, East and West Sussex County Councils, Sussex Health and Care Partnership and the NHS Sussex Commissioners, and thank them for the opportunity for past collaborations which lead to the conceiving of this manuscript. 

\textbf{Competing interests:} The authors have no relevant financial or non-financial interests to disclose.

\textbf{Author Contributions:} All authors contributed to the study conception and design. Material preparation, data collection and analysis were performed by JVY and RI. The draft of the manuscript was written by JVY and all authors commented on previous versions of the manuscript. All authors read and approved the final manuscript.

\bibliography{EJE_bib}


\begin{thebibliography}{67}
\ifx \bisbn   \undefined \def \bisbn  #1{ISBN #1}\fi
\ifx \binits  \undefined \def \binits#1{#1}\fi
\ifx \bauthor  \undefined \def \bauthor#1{#1}\fi
\ifx \batitle  \undefined \def \batitle#1{#1}\fi
\ifx \bjtitle  \undefined \def \bjtitle#1{#1}\fi
\ifx \bvolume  \undefined \def \bvolume#1{\textbf{#1}}\fi
\ifx \byear  \undefined \def \byear#1{#1}\fi
\ifx \bissue  \undefined \def \bissue#1{#1}\fi
\ifx \bfpage  \undefined \def \bfpage#1{#1}\fi
\ifx \blpage  \undefined \def \blpage #1{#1}\fi
\ifx \burl  \undefined \def \burl#1{\textsf{#1}}\fi
\ifx \doiurl  \undefined \def \doiurl#1{\url{https://doi.org/#1}}\fi
\ifx \betal  \undefined \def \betal{\textit{et al.}}\fi
\ifx \binstitute  \undefined \def \binstitute#1{#1}\fi
\ifx \binstitutionaled  \undefined \def \binstitutionaled#1{#1}\fi
\ifx \bctitle  \undefined \def \bctitle#1{#1}\fi
\ifx \beditor  \undefined \def \beditor#1{#1}\fi
\ifx \bpublisher  \undefined \def \bpublisher#1{#1}\fi
\ifx \bbtitle  \undefined \def \bbtitle#1{#1}\fi
\ifx \bedition  \undefined \def \bedition#1{#1}\fi
\ifx \bseriesno  \undefined \def \bseriesno#1{#1}\fi
\ifx \blocation  \undefined \def \blocation#1{#1}\fi
\ifx \bsertitle  \undefined \def \bsertitle#1{#1}\fi
\ifx \bsnm \undefined \def \bsnm#1{#1}\fi
\ifx \bsuffix \undefined \def \bsuffix#1{#1}\fi
\ifx \bparticle \undefined \def \bparticle#1{#1}\fi
\ifx \barticle \undefined \def \barticle#1{#1}\fi
\bibcommenthead
\ifx \bconfdate \undefined \def \bconfdate #1{#1}\fi
\ifx \botherref \undefined \def \botherref #1{#1}\fi
\ifx \url \undefined \def \url#1{\textsf{#1}}\fi
\ifx \bchapter \undefined \def \bchapter#1{#1}\fi
\ifx \bbook \undefined \def \bbook#1{#1}\fi
\ifx \bcomment \undefined \def \bcomment#1{#1}\fi
\ifx \oauthor \undefined \def \oauthor#1{#1}\fi
\ifx \citeauthoryear \undefined \def \citeauthoryear#1{#1}\fi
\ifx \endbibitem  \undefined \def \endbibitem {}\fi
\ifx \bconflocation  \undefined \def \bconflocation#1{#1}\fi
\ifx \arxivurl  \undefined \def \arxivurl#1{\textsf{#1}}\fi
\csname PreBibitemsHook\endcsname

\bibitem{DBJal20}
\begin{barticle}
\bauthor{\bsnm{Davies}, \binits{N.G.}},
\bauthor{\bsnm{Barnard}, \binits{R.C.}},
\bauthor{\bsnm{Jarvis}, \binits{C.I.}},
\bauthor{\bsnm{Russell}, \binits{T.W.}},
\bauthor{\bsnm{Semple}, \binits{M.G.}},
\bauthor{\bsnm{Jit}, \binits{M.}},
\bauthor{\bsnm{Edmunds}, \binits{W.J.}}, \betal:
\batitle{Association of tiered restrictions and a second lockdown with covid-19
  deaths and hospital admissions in england: a modelling study}.
\bjtitle{Lancet Infect. Dis.}
(\byear{2020}).
\doiurl{10.1016/S1473-3099(20)30984-1}
\end{barticle}
\endbibitem

\bibitem{CVAal20}
\begin{barticle}
\bauthor{\bsnm{Campillo-Funollet}, \binits{E.}},
\bauthor{\bsnm{Van~Yperen}, \binits{J.}},
\bauthor{\bsnm{Allman}, \binits{P.}},
\bauthor{\bsnm{Bell}, \binits{M.}},
\bauthor{\bsnm{Beresford}, \binits{W.}},
\bauthor{\bsnm{Clay}, \binits{J.}},
\bauthor{\bsnm{Dorey}, \binits{M.}},
\bauthor{\bsnm{Evans}, \binits{G.}},
\bauthor{\bsnm{Gilchrist}, \binits{K.}},
\bauthor{\bsnm{Memon}, \binits{A.}}, \betal:
\batitle{Predicting and forecasting the impact of local outbreaks of covid-19:
  Use of seir-d quantitative epidemiological modelling for healthcare demand
  and capacity}.
\bjtitle{Int. J. Epidemiol.}
\bvolume{50}(\bissue{4}),
\bfpage{1103}--\blpage{1113}
(\byear{2021}).
\doiurl{10.1093/ije/dyab106}
\end{barticle}
\endbibitem

\bibitem{DZSal20}
\begin{barticle}
\bauthor{\bsnm{Dehning}, \binits{J.}},
\bauthor{\bsnm{Zierenberg}, \binits{J.}},
\bauthor{\bsnm{Spitzner}, \binits{F.P.}},
\bauthor{\bsnm{Wibral}, \binits{M.}},
\bauthor{\bsnm{Neto}, \binits{J.P.}},
\bauthor{\bsnm{Wilczek}, \binits{M.}},
\bauthor{\bsnm{Priesemann}, \binits{V.}}:
\batitle{Inferring change points in the spread of covid-19 reveals the
  effectiveness of interventions}.
\bjtitle{Science}
(\byear{2020}).
\doiurl{10.1126/science.abb9789}
\end{barticle}
\endbibitem

\bibitem{EHMal20}
\begin{barticle}
\bauthor{\bsnm{Edejer}, \binits{T.T.-T.}},
\bauthor{\bsnm{Hanssen}, \binits{O.}},
\bauthor{\bsnm{Mirelman}, \binits{A.}},
\bauthor{\bsnm{Verboom}, \binits{P.}},
\bauthor{\bsnm{Lolong}, \binits{G.}},
\bauthor{\bsnm{Watson}, \binits{O.J.}},
\bauthor{\bsnm{Boulanger}, \binits{L.L.}},
\bauthor{\bsnm{Soucat}, \binits{A.}}:
\batitle{Projected health-care resource needs for an effective response to
  covid-19 in 73 low-income and middle-income countries: a modelling study}.
\bjtitle{Lancet Glob. Health}
(\byear{2020}).
\doiurl{10.1016/S2214-109X(20)30383-1}
\end{barticle}
\endbibitem

\bibitem{FLNal20}
\begin{botherref}
\oauthor{\bsnm{Ferguson}, \binits{N.M.}},
\oauthor{\bsnm{Laydon}, \binits{D.}},
\oauthor{\bsnm{Nedjati-Gilani}, \binits{G.}},
\oauthor{\bsnm{Imai}, \binits{N.}},
\oauthor{\bsnm{Ainslie}, \binits{K.}},
\oauthor{\bsnm{Baguelin}, \binits{M.}},
\oauthor{\bsnm{Bhatia}, \binits{S.}},
\oauthor{\bsnm{Boonyasiri}, \binits{A.}},
\oauthor{\bsnm{Cucunub{\'a}}, \binits{Z.}},
\oauthor{\bsnm{Cuomo-Dannenburg}, \binits{G.}}, et al.:
Impact of non-pharmaceutical interventions (npis) to reduce covid-19 mortality
  and healthcare demand
(2020).
\doiurl{10.25561/77482}
\end{botherref}
\endbibitem

\bibitem{FGLal20}
\begin{botherref}
\oauthor{\bsnm{Ferstad}, \binits{J.O.}},
\oauthor{\bsnm{Gu}, \binits{A.J.}},
\oauthor{\bsnm{Lee}, \binits{R.Y.}},
\oauthor{\bsnm{Thapa}, \binits{I.}},
\oauthor{\bsnm{Shin}, \binits{A.Y.}},
\oauthor{\bsnm{Salomon}, \binits{J.A.}},
\oauthor{\bsnm{Glynn}, \binits{P.}},
\oauthor{\bsnm{Shah}, \binits{N.H.}},
\oauthor{\bsnm{Milstein}, \binits{A.}},
\oauthor{\bsnm{Schulman}, \binits{K.}}, et al.:
A model to forecast regional demand for COVID-19 related hospital beds.
Preprint at \url{https://www.medrxiv.org/content/10.1101/2020.03.26.20044842v3}
(2020)
\end{botherref}
\endbibitem

\bibitem{GLBal20}
\begin{barticle}
\bauthor{\bsnm{Gariboldi}, \binits{M.I.}},
\bauthor{\bsnm{Lin}, \binits{V.}},
\bauthor{\bsnm{Bland}, \binits{J.}},
\bauthor{\bsnm{Auplish}, \binits{M.}},
\bauthor{\bsnm{Cawthorne}, \binits{A.}}:
\batitle{Foresight in the time of covid-19}.
\bjtitle{The Lancet Regional Health-Western Pacific}
\bvolume{6},
\bfpage{100049}
(\byear{2020}).
\doiurl{10.1016/j.lanwpc.2020.100049}
\end{barticle}
\endbibitem

\bibitem{JLJ20}
\begin{barticle}
\bauthor{\bsnm{Jewell}, \binits{N.P.}},
\bauthor{\bsnm{Lewnard}, \binits{J.A.}},
\bauthor{\bsnm{Jewell}, \binits{B.L.}}:
\batitle{Predictive mathematical models of the covid-19 pandemic: Underlying
  principles and value of projections}.
\bjtitle{JAMA}
\bvolume{323}(\bissue{19}),
\bfpage{1893}--\blpage{1894}
(\byear{2020}).
\doiurl{10.1001/jama.2020.6585}
\end{barticle}
\endbibitem

\bibitem{LTY20}
\begin{barticle}
\bauthor{\bsnm{Liu}, \binits{M.}},
\bauthor{\bsnm{Thomadsen}, \binits{R.}},
\bauthor{\bsnm{Yao}, \binits{S.}}:
\batitle{Forecasting the spread of covid-19 under different reopening
  strategies}.
\bjtitle{Sci. Rep.}
\bvolume{10}(\bissue{1}),
\bfpage{1}--\blpage{8}
(\byear{2020}).
\doiurl{10.1038/s41598-020-77292-8}
\end{barticle}
\endbibitem

\bibitem{LMKal20}
\begin{barticle}
\bauthor{\bsnm{Liu}, \binits{Y.}},
\bauthor{\bsnm{Morgenstern}, \binits{C.}},
\bauthor{\bsnm{Kelly}, \binits{J.}},
\bauthor{\bsnm{Lowe}, \binits{R.}},
\bauthor{\bsnm{Jit}, \binits{M.}}:
\batitle{The impact of non-pharmaceutical interventions on sars-cov-2
  transmission across 130 countries and territories}.
\bjtitle{BMC Med.}
\bvolume{19}(\bissue{1}),
\bfpage{1}--\blpage{12}
(\byear{2021}).
\doiurl{10.1186/s12916-020-01872-8}
\end{barticle}
\endbibitem

\bibitem{MG20}
\begin{barticle}
\bauthor{\bsnm{Mavragani}, \binits{A.}},
\bauthor{\bsnm{Gkillas}, \binits{K.}}:
\batitle{Covid-19 predictability in the united states using google trends time
  series}.
\bjtitle{Sci. Rep.}
\bvolume{10}(\bissue{1}),
\bfpage{1}--\blpage{12}
(\byear{2020}).
\doiurl{10.1038/s41598-020-77275-9}
\end{barticle}
\endbibitem

\bibitem{Ar20}
\begin{barticle}
\bauthor{\bsnm{Arino}, \binits{J.}}:
\batitle{Mathematical epidemiology in a data-rich world}.
\bjtitle{Infect. Dis. Model.}
\bvolume{5},
\bfpage{161}--\blpage{188}
(\byear{2020}).
\doiurl{10.1016/j.idm.2019.12.008}
\end{barticle}
\endbibitem

\bibitem{MHG21}
\begin{barticle}
\bauthor{\bsnm{Mercatelli}, \binits{D.}},
\bauthor{\bsnm{Holding}, \binits{A.N.}},
\bauthor{\bsnm{Giorgi}, \binits{F.M.}}:
\batitle{Web tools to fight pandemics: the covid-19 experience}.
\bjtitle{Brief. Bioinform.}
\bvolume{22}(\bissue{2}),
\bfpage{690}--\blpage{700}
(\byear{2021}).
\doiurl{10.1093/bib/bbaa261}
\end{barticle}
\endbibitem

\bibitem{goldacre}
\begin{botherref}
\oauthor{\bsnm{Goldacre}, \binits{B.}},
\oauthor{\bsnm{Morley}, \binits{J.}}:
Better, broader, safer: using health data for research and analysis.
A Review Commissioned by the Secretary of State for Health and Social Care
(2022)
\end{botherref}
\endbibitem

\bibitem{K05}
\begin{barticle}
\bauthor{\bsnm{Keeling}, \binits{M.J.}}:
\batitle{Models of foot-and-mouth disease}.
\bjtitle{Proc. R. Soc. B: Biol. Sci.}
\bvolume{272}(\bissue{1569}),
\bfpage{1195}--\blpage{1202}
(\byear{2005}).
\doiurl{10.1098/rspb.2004.3046}
\end{barticle}
\endbibitem

\bibitem{BVJFWE10}
\begin{barticle}
\bauthor{\bsnm{Baguelin}, \binits{M.}},
\bauthor{\bsnm{Van~Hoek}, \binits{A.J.}},
\bauthor{\bsnm{Jit}, \binits{M.}},
\bauthor{\bsnm{Flasche}, \binits{S.}},
\bauthor{\bsnm{White}, \binits{P.J.}},
\bauthor{\bsnm{Edmunds}, \binits{W.J.}}:
\batitle{Vaccination against pandemic influenza a/h1n1v in england: a real-time
  economic evaluation}.
\bjtitle{Vaccine}
\bvolume{28}(\bissue{12}),
\bfpage{2370}--\blpage{2384}
(\byear{2010}).
\doiurl{10.1016/j.vaccine.2010.01.002}
\end{barticle}
\endbibitem

\bibitem{FCFCCB06}
\begin{barticle}
\bauthor{\bsnm{Ferguson}, \binits{N.M.}},
\bauthor{\bsnm{Cummings}, \binits{D.A.}},
\bauthor{\bsnm{Fraser}, \binits{C.}},
\bauthor{\bsnm{Cajka}, \binits{J.C.}},
\bauthor{\bsnm{Cooley}, \binits{P.C.}},
\bauthor{\bsnm{Burke}, \binits{D.S.}}:
\batitle{Strategies for mitigating an influenza pandemic}.
\bjtitle{Nature}
\bvolume{442}(\bissue{7101}),
\bfpage{448}--\blpage{452}
(\byear{2006}).
\doiurl{10.1038/nature04795}
\end{barticle}
\endbibitem

\bibitem{AFG22}
\begin{barticle}
\bauthor{\bsnm{Avram}, \binits{F.}},
\bauthor{\bsnm{Freddi}, \binits{L.}},
\bauthor{\bsnm{Goreac}, \binits{D.}}:
\batitle{Optimal control of a sir epidemic with icu constraints and target
  objectives}.
\bjtitle{Appl. Math. Comput.}
\bvolume{418},
\bfpage{126816}
(\byear{2022}).
\doiurl{10.1016/j.amc.2021.126816}
\end{barticle}
\endbibitem

\bibitem{MSW20}
\begin{botherref}
\oauthor{\bsnm{Miclo}, \binits{L.}},
\oauthor{\bsnm{Spiro}, \binits{D.}},
\oauthor{\bsnm{Weibull}, \binits{J.}}:
Optimal epidemic suppression under an icu constraint: An analytical solution.
J. Math. Econ.,
102669
(2022).
\doiurl{10.1016/j.jmateco.2022.102669}
\end{botherref}
\endbibitem

\bibitem{DKM21}
\begin{barticle}
\bauthor{\bsnm{Di~Lauro}, \binits{F.}},
\bauthor{\bsnm{Kiss}, \binits{I.Z.}},
\bauthor{\bsnm{Miller}, \binits{J.C.}}:
\batitle{Optimal timing of one-shot interventions for epidemic control}.
\bjtitle{PLoS Comput. Biol.}
\bvolume{17}(\bissue{3}),
\bfpage{1008763}
(\byear{2021}).
\doiurl{10.1371/journal.pcbi.1008763}
\end{barticle}
\endbibitem

\bibitem{NPP16}
\begin{barticle}
\bauthor{\bsnm{Nowzari}, \binits{C.}},
\bauthor{\bsnm{Preciado}, \binits{V.M.}},
\bauthor{\bsnm{Pappas}, \binits{G.J.}}:
\batitle{Analysis and control of epidemics: A survey of spreading processes on
  complex networks}.
\bjtitle{IEEE Control Syst.}
\bvolume{36}(\bissue{1}),
\bfpage{26}--\blpage{46}
(\byear{2016}).
\doiurl{10.1109/MCS.2015.2495000}
\end{barticle}
\endbibitem

\bibitem{FIL21}
\begin{barticle}
\bauthor{\bsnm{Feng}, \binits{Y.}},
\bauthor{\bsnm{Iyer}, \binits{G.}},
\bauthor{\bsnm{Li}, \binits{L.}}:
\batitle{Scheduling fixed length quarantines to minimize the total number of
  fatalities during an epidemic}.
\bjtitle{J. Math. Biol.}
\bvolume{82}(\bissue{7}),
\bfpage{1}--\blpage{17}
(\byear{2021}).
\doiurl{10.1007/s00285-021-01615-0}
\end{barticle}
\endbibitem

\bibitem{JG22}
\begin{botherref}
\oauthor{\bsnm{Jana}, \binits{S.}},
\oauthor{\bsnm{Ghose}, \binits{D.}}:
Optimal Lockdown Management using Short Term COVID-19 Prediction Model.
Preprint at \url{https://arxiv.org/abs/2203.03488}
(2022)
\end{botherref}
\endbibitem

\bibitem{S21}
\begin{barticle}
\bauthor{\bsnm{Sontag}, \binits{E.D.}}:
\batitle{An explicit formula for minimizing the infected peak in an sir
  epidemic model when using a fixed number of complete lockdowns}.
\bjtitle{Int. J. Robust Nonlinear Control.}
(\byear{2021}).
\doiurl{10.1002/rnc.5701}
\end{barticle}
\endbibitem

\bibitem{DHR10}
\begin{barticle}
\bauthor{\bsnm{Diekmann}, \binits{O.}},
\bauthor{\bsnm{Heesterbeek}, \binits{J.}},
\bauthor{\bsnm{Roberts}, \binits{M.G.}}:
\batitle{The construction of next-generation matrices for compartmental
  epidemic models}.
\bjtitle{J. R. Soc. Interface}
\bvolume{7}(\bissue{47}),
\bfpage{873}--\blpage{885}
(\byear{2010}).
\doiurl{10.1098/rsif.2009.0386}
\end{barticle}
\endbibitem

\bibitem{CWVDM21}
\begin{botherref}
\oauthor{\bsnm{Campillo-Funollet}, \binits{E.}},
\oauthor{\bsnm{Wragg}, \binits{H.}},
\oauthor{\bsnm{Van~Yperen}, \binits{J.}},
\oauthor{\bsnm{Duong}, \binits{D.-L.}},
\oauthor{\bsnm{Madzvamuse}, \binits{A.}}:
Reformulating the SIR model in terms of the number of COVID-19 detected cases:
  well-posedness of the observational model.
Accepted in Philosophical Transactions A (Royal Society)
(2021)
\end{botherref}
\endbibitem

\bibitem{BG08}
\begin{bbook}
\bauthor{\bsnm{Butcher}, \binits{J.C.}}:
\bbtitle{Numerical Methods for Ordinary Differential Equations}.
\bpublisher{John Wiley \& Sons, Ltd},
\blocation{Chichester}
(\byear{2016})
\end{bbook}
\endbibitem

\bibitem{GH10}
\begin{bbook}
\bauthor{\bsnm{Griffiths}, \binits{D.F.}},
\bauthor{\bsnm{Higham}, \binits{D.J.}}:
\bbtitle{Numerical Methods for Ordinary Differential Equations: Initial Value
  Problems}.
\bpublisher{Springer},
\blocation{London}
(\byear{2010}).
\doiurl{10.1007/978-0-85729-148-6}
\end{bbook}
\endbibitem

\bibitem{P83}
\begin{barticle}
\bauthor{\bsnm{Petzold}, \binits{L.}}:
\batitle{Automatic selection of methods for solving stiff and nonstiff systems
  of ordinary differential equations}.
\bjtitle{SIAM J. Sci. Comput.}
\bvolume{4}(\bissue{1}),
\bfpage{136}--\blpage{148}
(\byear{1983}).
\doiurl{10.1137/0904010}
\end{barticle}
\endbibitem

\bibitem{K92}
\begin{barticle}
\bauthor{\bsnm{Knuth}, \binits{D.E.}}:
\batitle{Two notes on notation}.
\bjtitle{Am. Math. Mon.}
\bvolume{99}(\bissue{5}),
\bfpage{403}--\blpage{422}
(\byear{1992}).
\doiurl{10.1080/00029890.1992.11995869}
\end{barticle}
\endbibitem

\bibitem{MB20}
\begin{barticle}
\bauthor{\bsnm{Maier}, \binits{B.F.}},
\bauthor{\bsnm{Brockmann}, \binits{D.}}:
\batitle{Effective containment explains subexponential growth in recent
  confirmed covid-19 cases in china}.
\bjtitle{Science}
\bvolume{368}(\bissue{6492}),
\bfpage{742}--\blpage{746}
(\byear{2020}).
\doiurl{10.1126/science.abb4557}
\end{barticle}
\endbibitem

\bibitem{DD19}
\begin{barticle}
\bauthor{\bsnm{DeAngelis}, \binits{D.L.}},
\bauthor{\bsnm{Diaz}, \binits{S.G.}}:
\batitle{Decision-making in agent-based modeling: A current review and future
  prospectus}.
\bjtitle{Front. Ecol. Evol.}
\bvolume{6},
\bfpage{237}
(\byear{2019}).
\doiurl{10.3389/fevo.2018.00237}
\end{barticle}
\endbibitem

\bibitem{PSR98}
\begin{bchapter}
\bauthor{\bsnm{Parunak}, \binits{H.V.D.}},
\bauthor{\bsnm{Savit}, \binits{R.}},
\bauthor{\bsnm{Riolo}, \binits{R.L.}}:
\bctitle{Agent-based modeling vs. equation-based modeling: A case study and
  users’ guide}.
In: \beditor{\bsnm{Sichman}, \binits{J.S.}},
\beditor{\bsnm{Conte}, \binits{R.}},
\beditor{\bsnm{Gilbert}, \binits{N.}} (eds.)
\bbtitle{International Workshop on Multi-agent Systems and Agent-based
  Simulation},
pp. \bfpage{10}--\blpage{25}.
\bpublisher{Springer},
\blocation{Berlin}
(\byear{1998}).
\doiurl{10.1007/10692956_2}
\end{bchapter}
\endbibitem

\bibitem{RG19}
\begin{bbook}
\bauthor{\bsnm{Railsback}, \binits{S.F.}},
\bauthor{\bsnm{Grimm}, \binits{V.}}:
\bbtitle{Agent-based and Individual-based Modeling: a Practical Introduction}.
\bpublisher{Princeton university press},
\blocation{Princeton}
(\byear{2019})
\end{bbook}
\endbibitem

\bibitem{SJNSS18}
\begin{botherref}
\oauthor{\bsnm{Sheikh}, \binits{M.J.}},
\oauthor{\bsnm{Jandaghi}, \binits{G.}},
\oauthor{\bsnm{Naeini}, \binits{A.B.}},
\oauthor{\bsnm{Shafia}, \binits{M.A.}},
\oauthor{\bsnm{Sabzian}, \binits{H.}}:
A review of agent-based modeling (abm) concepts and some of its main
  applications in management science.
Iran. J. Manag. Stud.
\textbf{11}(4)
(2018).
\doiurl{10.22059/IJMS.2018.261178.673190}
\end{botherref}
\endbibitem

\bibitem{D05}
\begin{botherref}
\oauthor{\bsnm{Dunham}, \binits{J.B.}}:
An agent-based spatially explicit epidemiological model in mason.
J. Artif. Soc. Soc. Simul.
\textbf{9}(1)
(2005)
\end{botherref}
\endbibitem

\bibitem{LS12}
\begin{barticle}
\bauthor{\bsnm{Luke}, \binits{D.A.}},
\bauthor{\bsnm{Stamatakis}, \binits{K.A.}}:
\batitle{Systems science methods in public health: dynamics, networks, and
  agents}.
\bjtitle{Annu. Rev. Public Health}
\bvolume{33},
\bfpage{357}--\blpage{376}
(\byear{2012}).
\doiurl{10.1146/annurev-publhealth-031210-101222}
\end{barticle}
\endbibitem

\bibitem{PD09}
\begin{barticle}
\bauthor{\bsnm{Perez}, \binits{L.}},
\bauthor{\bsnm{Dragicevic}, \binits{S.}}:
\batitle{An agent-based approach for modeling dynamics of contagious disease
  spread}.
\bjtitle{Int. J. Health Geogr.}
\bvolume{8}(\bissue{1}),
\bfpage{1}--\blpage{17}
(\byear{2009}).
\doiurl{10.1186/1476-072X-8-50}
\end{barticle}
\endbibitem

\bibitem{FCCFRMIB05}
\begin{barticle}
\bauthor{\bsnm{Ferguson}, \binits{N.M.}},
\bauthor{\bsnm{Cummings}, \binits{D.A.}},
\bauthor{\bsnm{Cauchemez}, \binits{S.}},
\bauthor{\bsnm{Fraser}, \binits{C.}},
\bauthor{\bsnm{Riley}, \binits{S.}},
\bauthor{\bsnm{Meeyai}, \binits{A.}},
\bauthor{\bsnm{Iamsirithaworn}, \binits{S.}},
\bauthor{\bsnm{Burke}, \binits{D.S.}}:
\batitle{Strategies for containing an emerging influenza pandemic in southeast
  asia}.
\bjtitle{Nature}
\bvolume{437}(\bissue{7056}),
\bfpage{209}--\blpage{214}
(\byear{2005}).
\doiurl{10.1038/nature04017}
\end{barticle}
\endbibitem

\bibitem{HFELCLXFVAGal08}
\begin{barticle}
\bauthor{\bsnm{Halloran}, \binits{M.E.}},
\bauthor{\bsnm{Ferguson}, \binits{N.M.}},
\bauthor{\bsnm{Eubank}, \binits{S.}},
\bauthor{\bsnm{Longini}, \binits{I.M.}},
\bauthor{\bsnm{Cummings}, \binits{D.A.}},
\bauthor{\bsnm{Lewis}, \binits{B.}},
\bauthor{\bsnm{Xu}, \binits{S.}},
\bauthor{\bsnm{Fraser}, \binits{C.}},
\bauthor{\bsnm{Vullikanti}, \binits{A.}},
\bauthor{\bsnm{Germann}, \binits{T.C.}}, \betal:
\batitle{Modeling targeted layered containment of an influenza pandemic in the
  united states}.
\bjtitle{PNAS}
\bvolume{105}(\bissue{12}),
\bfpage{4639}--\blpage{4644}
(\byear{2008}).
\doiurl{10.1073/pnas.0706849105}
\end{barticle}
\endbibitem

\bibitem{AHWLPWESRD21}
\begin{barticle}
\bauthor{\bsnm{Abueg}, \binits{M.}},
\bauthor{\bsnm{Hinch}, \binits{R.}},
\bauthor{\bsnm{Wu}, \binits{N.}},
\bauthor{\bsnm{Liu}, \binits{L.}},
\bauthor{\bsnm{Probert}, \binits{W.}},
\bauthor{\bsnm{Wu}, \binits{A.}},
\bauthor{\bsnm{Eastham}, \binits{P.}},
\bauthor{\bsnm{Shafi}, \binits{Y.}},
\bauthor{\bsnm{Rosencrantz}, \binits{M.}},
\bauthor{\bsnm{Dikovsky}, \binits{M.}}, \betal:
\batitle{Modeling the effect of exposure notification and non-pharmaceutical
  interventions on covid-19 transmission in washington state}.
\bjtitle{NPJ Digit. Med.}
\bvolume{4}(\bissue{1}),
\bfpage{1}--\blpage{10}
(\byear{2021}).
\doiurl{10.1038/s41746-021-00422-7}
\end{barticle}
\endbibitem

\bibitem{HPNKWHLBZS21}
\begin{barticle}
\bauthor{\bsnm{Hinch}, \binits{R.}},
\bauthor{\bsnm{Probert}, \binits{W.J.}},
\bauthor{\bsnm{Nurtay}, \binits{A.}},
\bauthor{\bsnm{Kendall}, \binits{M.}},
\bauthor{\bsnm{Wymant}, \binits{C.}},
\bauthor{\bsnm{Hall}, \binits{M.}},
\bauthor{\bsnm{Lythgoe}, \binits{K.}},
\bauthor{\bsnm{Bulas~Cruz}, \binits{A.}},
\bauthor{\bsnm{Zhao}, \binits{L.}},
\bauthor{\bsnm{Stewart}, \binits{A.}}, \betal:
\batitle{Openabm-covid19—an agent-based model for non-pharmaceutical
  interventions against covid-19 including contact tracing}.
\bjtitle{PLoS Comput. Biol.}
\bvolume{17}(\bissue{7}),
\bfpage{1009146}
(\byear{2021}).
\doiurl{10.1371/journal.pcbi.1009146}
\end{barticle}
\endbibitem

\bibitem{KSMARHNRCSH21}
\begin{barticle}
\bauthor{\bsnm{Kerr}, \binits{C.C.}},
\bauthor{\bsnm{Stuart}, \binits{R.M.}},
\bauthor{\bsnm{Mistry}, \binits{D.}},
\bauthor{\bsnm{Abeysuriya}, \binits{R.G.}},
\bauthor{\bsnm{Rosenfeld}, \binits{K.}},
\bauthor{\bsnm{Hart}, \binits{G.R.}},
\bauthor{\bsnm{N{\'u}{\~n}ez}, \binits{R.C.}},
\bauthor{\bsnm{Cohen}, \binits{J.A.}},
\bauthor{\bsnm{Selvaraj}, \binits{P.}},
\bauthor{\bsnm{Hagedorn}, \binits{B.}}, \betal:
\batitle{Covasim: an agent-based model of covid-19 dynamics and interventions}.
\bjtitle{PLOS Comput. Biol.}
\bvolume{17}(\bissue{7}),
\bfpage{1009149}
(\byear{2021}).
\doiurl{10.1371/journal.pcbi.1009149}
\end{barticle}
\endbibitem

\bibitem{PKSMKVB20}
\begin{barticle}
\bauthor{\bsnm{Panovska-Griffiths}, \binits{J.}},
\bauthor{\bsnm{Kerr}, \binits{C.C.}},
\bauthor{\bsnm{Stuart}, \binits{R.M.}},
\bauthor{\bsnm{Mistry}, \binits{D.}},
\bauthor{\bsnm{Klein}, \binits{D.J.}},
\bauthor{\bsnm{Viner}, \binits{R.M.}},
\bauthor{\bsnm{Bonell}, \binits{C.}}:
\batitle{Determining the optimal strategy for reopening schools, the impact of
  test and trace interventions, and the risk of occurrence of a second covid-19
  epidemic wave in the uk: a modelling study}.
\bjtitle{Lancet Child Adolesc. Health}
\bvolume{4}(\bissue{11}),
\bfpage{817}--\blpage{827}
(\byear{2020}).
\doiurl{10.1016/S2352-4642(20)30250-9}
\end{barticle}
\endbibitem

\bibitem{CDE18}
\begin{barticle}
\bauthor{\bsnm{Champredon}, \binits{D.}},
\bauthor{\bsnm{Dushoff}, \binits{J.}},
\bauthor{\bsnm{Earn}, \binits{D.J.}}:
\batitle{Equivalence of the erlang-distributed seir epidemic model and the
  renewal equation}.
\bjtitle{SIAM J. Appl. Math.}
\bvolume{78}(\bissue{6}),
\bfpage{3258}--\blpage{3278}
(\byear{2018}).
\doiurl{10.1137/18M1186411}
\end{barticle}
\endbibitem

\bibitem{AG12}
\begin{bbook}
\bauthor{\bsnm{Allgower}, \binits{E.L.}},
\bauthor{\bsnm{Georg}, \binits{K.}}:
\bbtitle{Numerical Continuation Methods: an Introduction}.
\bpublisher{Springer},
\blocation{Berlin}
(\byear{2012}).
\doiurl{10.1007/978-3-642-61257-2}
\end{bbook}
\endbibitem

\bibitem{B81}
\begin{bchapter}
\bauthor{\bsnm{Bock}, \binits{H.G.}}:
\bctitle{Numerical treatment of inverse problems in chemical reaction
  kinetics}.
In: \beditor{\bsnm{Ebert}, \binits{K.H.}},
\beditor{\bsnm{Deufflhard}, \binits{P.}},
\beditor{\bsnm{J\"{a}ger}, \binits{W.}} (eds.)
\bbtitle{Modelling of Chemical Reaction Systems},
pp. \bfpage{102}--\blpage{125}.
\bpublisher{Springer},
\blocation{Berlin}
(\byear{1981}).
\doiurl{10.1007/978-3-642-68220-9}
\end{bchapter}
\endbibitem

\bibitem{HMTMW00}
\begin{bchapter}
\bauthor{\bsnm{Horbelt}, \binits{W.}},
\bauthor{\bsnm{M{\"u}ller}, \binits{T.}},
\bauthor{\bsnm{Timmer}, \binits{J.}},
\bauthor{\bsnm{Melzer}, \binits{W.}},
\bauthor{\bsnm{Winkler}, \binits{K.}}:
\bctitle{Analysis of nonlinear differential equations: parameter estimation and
  model selection}.
In: \beditor{\bsnm{Brause}, \binits{R.W.}},
\beditor{\bsnm{Hanisch}, \binits{E.}} (eds.)
\bbtitle{International Symposium on Medical Data Analysis},
pp. \bfpage{152}--\blpage{159}.
\bpublisher{Springer},
\blocation{Berlin}
(\byear{2000}).
\doiurl{10.1007/3-540-39949-6_19}
\end{bchapter}
\endbibitem

\bibitem{RHCC07}
\begin{barticle}
\bauthor{\bsnm{Ramsay}, \binits{J.O.}},
\bauthor{\bsnm{Hooker}, \binits{G.}},
\bauthor{\bsnm{Campbell}, \binits{D.}},
\bauthor{\bsnm{Cao}, \binits{J.}}:
\batitle{Parameter estimation for differential equations: a generalized
  smoothing approach}.
\bjtitle{J. R. Stat. Soc. Series B Stat. Methodol.}
\bvolume{69}(\bissue{5}),
\bfpage{741}--\blpage{796}
(\byear{2007}).
\doiurl{10.1111/j.1467-9868.2007.00610.x}
\end{barticle}
\endbibitem

\bibitem{BI20}
\begin{barticle}
\bauthor{\bsnm{Bar-Zeev}, \binits{N.}},
\bauthor{\bsnm{Inglesby}, \binits{T.}}:
\batitle{Covid-19 vaccines: early success and remaining challenges}.
\bjtitle{Lancet}
(\byear{2020}).
\doiurl{10.1016/S0140-6736(20)31867-5}
\end{barticle}
\endbibitem

\bibitem{BRKLCGL21}
\begin{barticle}
\bauthor{\bsnm{Bubar}, \binits{K.M.}},
\bauthor{\bsnm{Reinholt}, \binits{K.}},
\bauthor{\bsnm{Kissler}, \binits{S.M.}},
\bauthor{\bsnm{Lipsitch}, \binits{M.}},
\bauthor{\bsnm{Cobey}, \binits{S.}},
\bauthor{\bsnm{Grad}, \binits{Y.H.}},
\bauthor{\bsnm{Larremore}, \binits{D.B.}}:
\batitle{Model-informed covid-19 vaccine prioritization strategies by age and
  serostatus}.
\bjtitle{Science}
\bvolume{371}(\bissue{6532}),
\bfpage{916}--\blpage{921}
(\byear{2021}).
\doiurl{10.1126/science.abe6959}
\end{barticle}
\endbibitem

\bibitem{LSHH20}
\begin{barticle}
\bauthor{\bsnm{Lurie}, \binits{N.}},
\bauthor{\bsnm{Saville}, \binits{M.}},
\bauthor{\bsnm{Hatchett}, \binits{R.}},
\bauthor{\bsnm{Halton}, \binits{J.}}:
\batitle{Developing covid-19 vaccines at pandemic speed}.
\bjtitle{N. Engl. J. Med.}
\bvolume{382}(\bissue{21}),
\bfpage{1969}--\blpage{1973}
(\byear{2020}).
\doiurl{10.1056/NEJMp2005630}
\end{barticle}
\endbibitem

\bibitem{MHTDK21}
\begin{barticle}
\bauthor{\bsnm{Moore}, \binits{S.}},
\bauthor{\bsnm{Hill}, \binits{E.M.}},
\bauthor{\bsnm{Tildesley}, \binits{M.J.}},
\bauthor{\bsnm{Dyson}, \binits{L.}},
\bauthor{\bsnm{Keeling}, \binits{M.J.}}:
\batitle{Vaccination and non-pharmaceutical interventions for covid-19: a
  mathematical modelling study}.
\bjtitle{Lancet Infect. Dis.}
\bvolume{21}(\bissue{6}),
\bfpage{793}--\blpage{802}
(\byear{2021}).
\doiurl{10.1016/S1473-3099(21)00143-2}
\end{barticle}
\endbibitem

\bibitem{SWBal20}
\begin{barticle}
\bauthor{\bsnm{Saad-Roy}, \binits{C.M.}},
\bauthor{\bsnm{Wagner}, \binits{C.E.}},
\bauthor{\bsnm{Baker}, \binits{R.E.}},
\bauthor{\bsnm{Morris}, \binits{S.E.}},
\bauthor{\bsnm{Farrar}, \binits{J.}},
\bauthor{\bsnm{Graham}, \binits{A.L.}},
\bauthor{\bsnm{Levin}, \binits{S.A.}},
\bauthor{\bsnm{Mina}, \binits{M.J.}},
\bauthor{\bsnm{Metcalf}, \binits{C.J.E.}},
\bauthor{\bsnm{Grenfell}, \binits{B.T.}}:
\batitle{Immune life history, vaccination, and the dynamics of sars-cov-2 over
  the next 5 years}.
\bjtitle{Science}
\bvolume{370}(\bissue{6518}),
\bfpage{811}--\blpage{818}
(\byear{2020}).
\doiurl{10.1126/science.abd7343}
\end{barticle}
\endbibitem

\bibitem{GTNT07}
\begin{barticle}
\bauthor{\bsnm{Gao}, \binits{S.}},
\bauthor{\bsnm{Teng}, \binits{Z.}},
\bauthor{\bsnm{Nieto}, \binits{J.J.}},
\bauthor{\bsnm{Torres}, \binits{A.}}:
\batitle{Analysis of an sir epidemic model with pulse vaccination and
  distributed time delay}.
\bjtitle{J. Biotechnol. Biomed.}
(\byear{2007}).
\doiurl{10.1155/2007/64870}
\end{barticle}
\endbibitem

\bibitem{DBDal20}
\begin{botherref}
\oauthor{\bsnm{Di~Lauro}, \binits{F.}},
\oauthor{\bsnm{Berthouze}, \binits{L.}},
\oauthor{\bsnm{Dorey}, \binits{M.D.}},
\oauthor{\bsnm{Miller}, \binits{J.C.}},
\oauthor{\bsnm{Kiss}, \binits{I.Z.}}:
The impact of network properties and mixing on control measures and
  disease-induced herd immunity in epidemic models: a mean-field model
  perspective.
Preprint at \url{https://arxiv.org/abs/2007.06975}
(2020)
\end{botherref}
\endbibitem

\bibitem{KKCal20}
\begin{botherref}
\oauthor{\bsnm{Klepac}, \binits{P.}},
\oauthor{\bsnm{Kucharski}, \binits{A.J.}},
\oauthor{\bsnm{Conlan}, \binits{A.J.}},
\oauthor{\bsnm{Kissler}, \binits{S.}},
\oauthor{\bsnm{Tang}, \binits{M.}},
\oauthor{\bsnm{Fry}, \binits{H.}},
\oauthor{\bsnm{Gog}, \binits{J.R.}}:
Contacts in context: large-scale setting-specific social mixing matrices from
  the BBC Pandemic project.
Preprint at \url{https://www.medrxiv.org/content/10.1101/2020.02.16.20023754v2}
(2020)
\end{botherref}
\endbibitem

\bibitem{NMDal20}
\begin{barticle}
\bauthor{\bsnm{Nussbaumer-Streit}, \binits{B.}},
\bauthor{\bsnm{Mayr}, \binits{V.}},
\bauthor{\bsnm{Dobrescu}, \binits{A.I.}},
\bauthor{\bsnm{Chapman}, \binits{A.}},
\bauthor{\bsnm{Persad}, \binits{E.}},
\bauthor{\bsnm{Klerings}, \binits{I.}},
\bauthor{\bsnm{Wagner}, \binits{G.}},
\bauthor{\bsnm{Siebert}, \binits{U.}},
\bauthor{\bsnm{Christof}, \binits{C.}},
\bauthor{\bsnm{Zachariah}, \binits{C.}}, \betal:
\batitle{Quarantine alone or in combination with other public health measures
  to control covid-19: a rapid review}.
\bjtitle{Cochrane Database Syst. Rev.}
(\bissue{4})
(\byear{2020}).
\doiurl{10.1002/14651858.CD013574.pub2}
\end{barticle}
\endbibitem

\bibitem{CBLal20}
\begin{barticle}
\bauthor{\bsnm{Carf{\`\i}}, \binits{A.}},
\bauthor{\bsnm{Bernabei}, \binits{R.}},
\bauthor{\bsnm{Landi}, \binits{F.}}, \betal:
\batitle{Persistent symptoms in patients after acute covid-19}.
\bjtitle{JAMA}
\bvolume{324}(\bissue{6}),
\bfpage{603}--\blpage{605}
(\byear{2020}).
\doiurl{10.1001/jama.2020.12603}
\end{barticle}
\endbibitem

\bibitem{YWVal20}
\begin{barticle}
\bauthor{\bsnm{Yelin}, \binits{D.}},
\bauthor{\bsnm{Wirtheim}, \binits{E.}},
\bauthor{\bsnm{Vetter}, \binits{P.}},
\bauthor{\bsnm{Kalil}, \binits{A.C.}},
\bauthor{\bsnm{Bruchfeld}, \binits{J.}},
\bauthor{\bsnm{Runold}, \binits{M.}},
\bauthor{\bsnm{Guaraldi}, \binits{G.}},
\bauthor{\bsnm{Mussini}, \binits{C.}},
\bauthor{\bsnm{Gudiol}, \binits{C.}},
\bauthor{\bsnm{Pujol}, \binits{M.}}, \betal:
\batitle{Long-term consequences of covid-19: research needs}.
\bjtitle{Lancet Infect. Dis.}
(\byear{2020}).
\doiurl{10.1016/S1473-3099(20)30701-5}
\end{barticle}
\endbibitem

\bibitem{Al20}
\begin{botherref}
\oauthor{\bsnm{Alwan}, \binits{N.A.}}:
A negative covid-19 test does not mean recovery.
Nature,
170--170
(2020).
\doiurl{10.1038/d41586-020-02335-z}
\end{botherref}
\endbibitem

\bibitem{Sh20}
\begin{barticle}
\bauthor{\bsnm{Sheehy}, \binits{L.M.}}:
\batitle{Considerations for postacute rehabilitation for survivors of
  covid-19}.
\bjtitle{JMIR public health and surveillance}
\bvolume{6}(\bissue{2}),
\bfpage{19462}
(\byear{2020}).
\doiurl{10.2196/19462}
\end{barticle}
\endbibitem

\bibitem{SMVGPBPCKAC21}
\begin{barticle}
\bauthor{\bsnm{Sudre}, \binits{C.H.}},
\bauthor{\bsnm{Murray}, \binits{B.}},
\bauthor{\bsnm{Varsavsky}, \binits{T.}},
\bauthor{\bsnm{Graham}, \binits{M.S.}},
\bauthor{\bsnm{Penfold}, \binits{R.S.}},
\bauthor{\bsnm{Bowyer}, \binits{R.C.}},
\bauthor{\bsnm{Pujol}, \binits{J.C.}},
\bauthor{\bsnm{Klaser}, \binits{K.}},
\bauthor{\bsnm{Antonelli}, \binits{M.}},
\bauthor{\bsnm{Canas}, \binits{L.S.}}, \betal:
\batitle{Attributes and predictors of long covid}.
\bjtitle{Nat. Med.}
\bvolume{27}(\bissue{4}),
\bfpage{626}--\blpage{631}
(\byear{2021}).
\doiurl{10.1038/s41591-021-01292-y}
\end{barticle}
\endbibitem

\bibitem{numpy}
\begin{barticle}
\bauthor{\bsnm{Harris}, \binits{C.R.}},
\bauthor{\bsnm{Millman}, \binits{K.J.}},
\bauthor{\bparticle{van~der} \bsnm{Walt}, \binits{S.J.}},
\bauthor{\bsnm{Gommers}, \binits{R.}},
\bauthor{\bsnm{Virtanen}, \binits{P.}},
\bauthor{\bsnm{Cournapeau}, \binits{D.}},
\bauthor{\bsnm{Wieser}, \binits{E.}},
\bauthor{\bsnm{Taylor}, \binits{J.}},
\bauthor{\bsnm{Berg}, \binits{S.}},
\bauthor{\bsnm{Smith}, \binits{N.J.}},
\bauthor{\bsnm{Kern}, \binits{R.}},
\bauthor{\bsnm{Picus}, \binits{M.}},
\bauthor{\bsnm{Hoyer}, \binits{S.}},
\bauthor{\bparticle{van} \bsnm{Kerkwijk}, \binits{M.H.}},
\bauthor{\bsnm{Brett}, \binits{M.}},
\bauthor{\bsnm{Haldane}, \binits{A.}},
\bauthor{\bparticle{Fernández~del} \bsnm{Río}, \binits{J.}},
\bauthor{\bsnm{Wiebe}, \binits{M.}},
\bauthor{\bsnm{Peterson}, \binits{P.}},
\bauthor{\bsnm{Gérard-Marchant}, \binits{P.}},
\bauthor{\bsnm{Sheppard}, \binits{K.}},
\bauthor{\bsnm{Reddy}, \binits{T.}},
\bauthor{\bsnm{Weckesser}, \binits{W.}},
\bauthor{\bsnm{Abbasi}, \binits{H.}},
\bauthor{\bsnm{Gohlke}, \binits{C.}},
\bauthor{\bsnm{Oliphant}, \binits{T.E.}}:
\batitle{Array programming with {NumPy}}.
\bjtitle{Nature}
\bvolume{585},
\bfpage{357}--\blpage{362}
(\byear{2020}).
\doiurl{10.1038/s41586-020-2649-2}
\end{barticle}
\endbibitem

\bibitem{matplotlib}
\begin{barticle}
\bauthor{\bsnm{Hunter}, \binits{J.D.}}:
\batitle{Matplotlib: A 2d graphics environment}.
\bjtitle{Comput. Sci. Eng.}
\bvolume{9}(\bissue{03}),
\bfpage{90}--\blpage{95}
(\byear{2007}).
\doiurl{10.1109/MCSE.2007.55}
\end{barticle}
\endbibitem

\bibitem{pandas}
\begin{bchapter}
\bauthor{\bsnm{McKinney}, \binits{W.}}, \betal:
\bctitle{Data structures for statistical computing in python}.
In: \beditor{\bparticle{van~der} \bsnm{Walt}, \binits{S.}},
\beditor{\bsnm{Millman}, \binits{J.}} (eds.)
\bbtitle{Proceedings of the 9th Python in Science Conference},
vol. \bseriesno{445},
pp. \bfpage{51}--\blpage{56}
(\byear{2010}).
\doiurl{10.25080/Majora-92bf1922-00a}
\end{bchapter}
\endbibitem

\bibitem{scipy}
\begin{barticle}
\bauthor{\bsnm{Virtanen}, \binits{P.}},
\bauthor{\bsnm{Gommers}, \binits{R.}},
\bauthor{\bsnm{Oliphant}, \binits{T.E.}},
\bauthor{\bsnm{Haberland}, \binits{M.}},
\bauthor{\bsnm{Reddy}, \binits{T.}},
\bauthor{\bsnm{Cournapeau}, \binits{D.}},
\bauthor{\bsnm{Burovski}, \binits{E.}},
\bauthor{\bsnm{Peterson}, \binits{P.}},
\bauthor{\bsnm{Weckesser}, \binits{W.}},
\bauthor{\bsnm{Bright}, \binits{J.}},
\bauthor{\bsnm{{van der Walt}}, \binits{S.J.}},
\bauthor{\bsnm{Brett}, \binits{M.}},
\bauthor{\bsnm{Wilson}, \binits{J.}},
\bauthor{\bsnm{Millman}, \binits{K.J.}},
\bauthor{\bsnm{Mayorov}, \binits{N.}},
\bauthor{\bsnm{Nelson}, \binits{A.R.J.}},
\bauthor{\bsnm{Jones}, \binits{E.}},
\bauthor{\bsnm{Kern}, \binits{R.}},
\bauthor{\bsnm{Larson}, \binits{E.}},
\bauthor{\bsnm{Carey}, \binits{C.J.}},
\bauthor{\bsnm{Polat}, \binits{{\. I}.}},
\bauthor{\bsnm{Feng}, \binits{Y.}},
\bauthor{\bsnm{Moore}, \binits{E.W.}},
\bauthor{\bsnm{{VanderPlas}}, \binits{J.}},
\bauthor{\bsnm{Laxalde}, \binits{D.}},
\bauthor{\bsnm{Perktold}, \binits{J.}},
\bauthor{\bsnm{Cimrman}, \binits{R.}},
\bauthor{\bsnm{Henriksen}, \binits{I.}},
\bauthor{\bsnm{Quintero}, \binits{E.A.}},
\bauthor{\bsnm{Harris}, \binits{C.R.}},
\bauthor{\bsnm{Archibald}, \binits{A.M.}},
\bauthor{\bsnm{Ribeiro}, \binits{A.H.}},
\bauthor{\bsnm{Pedregosa}, \binits{F.}},
\bauthor{\bsnm{{van Mulbregt}}, \binits{P.}},
\bauthor{\bsnm{{SciPy 1.0 Contributors}}}:
\batitle{{{SciPy} 1.0: Fundamental Algorithms for Scientific Computing in
  Python}}.
\bjtitle{Nat. Methods}
\bvolume{17},
\bfpage{261}--\blpage{272}
(\byear{2020}).
\doiurl{10.1038/s41592-019-0686-2}
\end{barticle}
\endbibitem

\end{thebibliography}

\appendix
\renewcommand{\thesection}{Appendix \Alph{section}}
\section{Verification of parameters}
\label{justification} 

In order to verify that the agent-based approach was modelling the exact same problem as the SEIR-D model (and thus the data) without having to do any parameter estimation, we chose to check the following criteria:
\begin{itemize}
    \item is the mean of the output of the agent-based approach almost indistinguishable to the output of the SEIR-D model?
    \item is the mean of the time spent in the states $E$, $U$ and $I$ close to the inverse of the associated rate parameter? 
    \item is the proportion of people going into $U$, $D_U$ and $D_H$ close to the associated probability parameters?
\end{itemize}
The first criterion was measured by plotting both approaches, the second criterion was calculated by measuring the frequency of the length of stay of each agent in each compartment during the simulation, and the third criterion was calculated by counting the number of agents who went along each decision. 

\begin{figure}[!htb]
    \centering
    \includegraphics[width=0.75\textwidth]{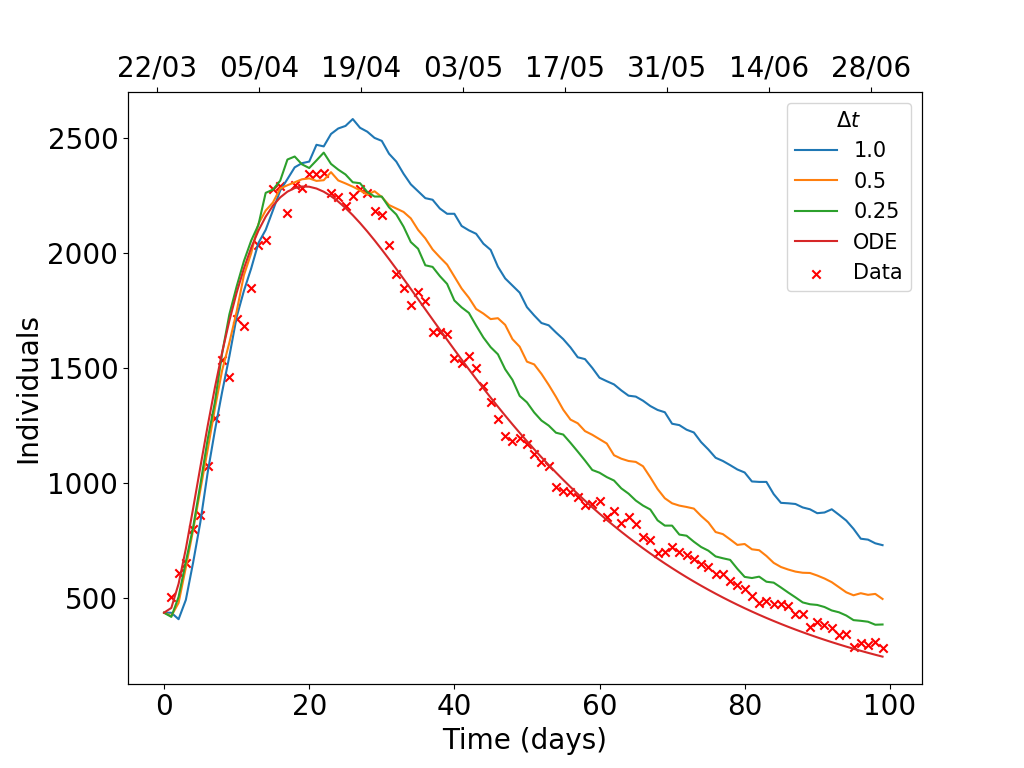}
    \caption{Comparison of the agent-based approach to the SEIR-D model, reducing $\dt$ and considering beds occupied for the South East without using the correction term.}
    \label{fig:no_correction_conv}
\end{figure}

\begin{figure}[!htb]
    \centering
    \includegraphics[width=0.75\textwidth]{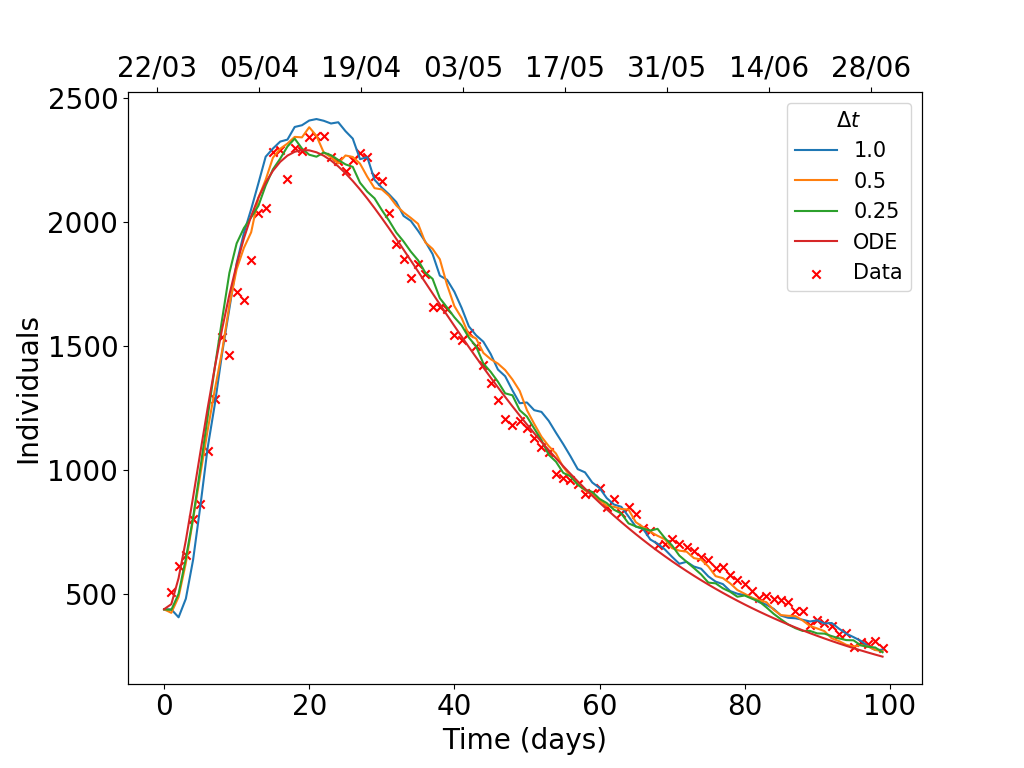}
    \caption{Comparison of the agent-based approach to the SEIR-D model, reducing $\dt$ and considering beds occupied for the South East using the correction term.}
    \label{fig:correction_conv}
\end{figure}

\begin{figure}[!htb]
    \centering
    \includegraphics[width=\textwidth]{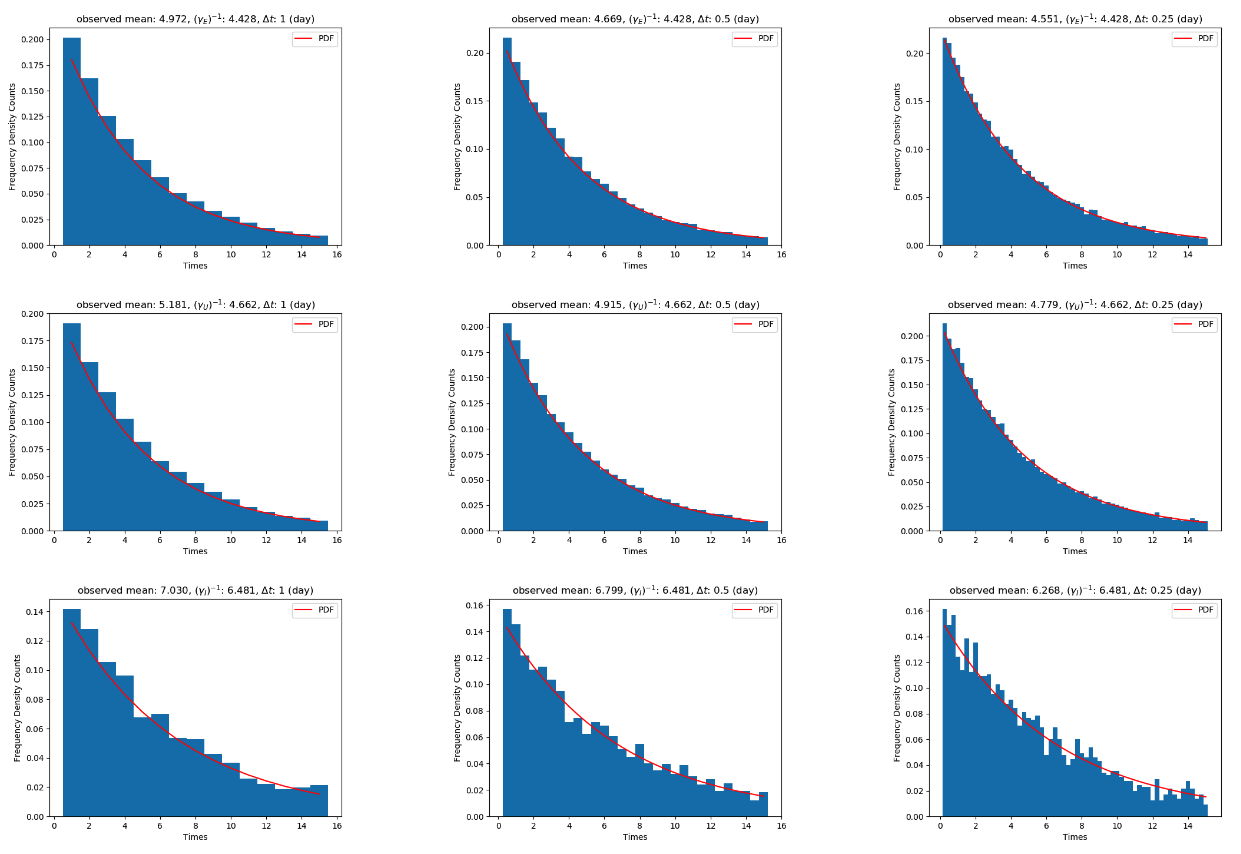}
    \caption{Frequency density of lengths of agents in certain state without the use of the correction term. The titles of each plot depict the observed mean of the frequencies of the lengths, the expected mean given the state and fitted parameter, and the unit of days being considered. The first row corresponds to lengths of time agents spend in the $E$ state, which should correspond to an exponential distribution with mean $\gamma_E$, the second row corresponds to lengths of time agents spend in the $U$ state, which should correspond to an exponential distribution with mean $\gamma_U$, and the final row corresponds to lengths of time agents spend in the $I$ state, which should correspond to an exponential distribution with mean $\gamma_I$. The first column uses a time unit of 1 day, the second column uses a time unit half a day and the last column uses a time unit of a quarter of a day.}
    \label{fig:no_correction}
\end{figure}

\begin{figure}[!htb]
    \centering
    \includegraphics[width=\textwidth]{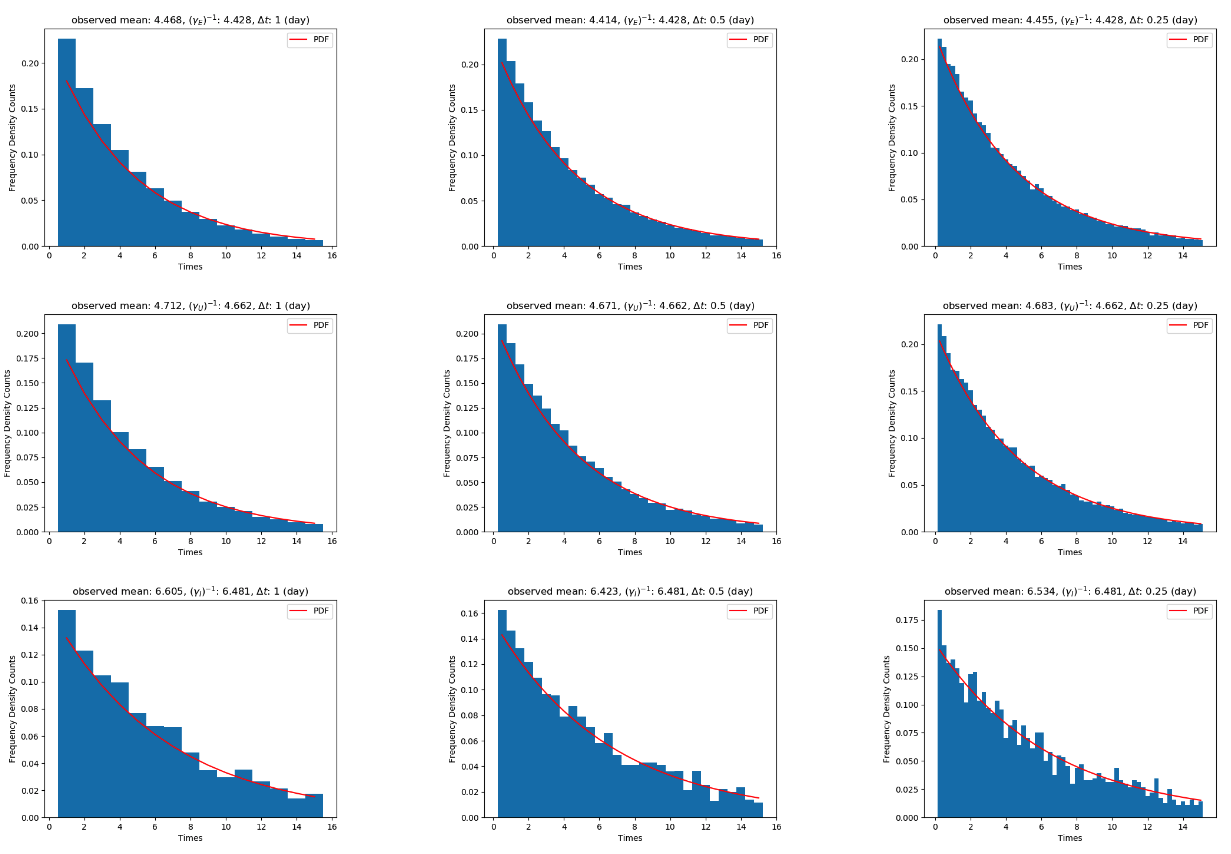}
    \caption{Frequency density of lengths of agents in certain state with the use of the correction term. The titles of each plot depict the observed mean of the frequencies of the lengths, the expected mean given the state and fitted parameter, and the unit of days being considered. The first row corresponds to lengths of time agents spend in the $E$ state, which does correspond to an exponential distribution with mean $\gamma_E$, the second row corresponds to lengths of time agents spend in the $U$ state, which does correspond to an exponential distribution with mean $\gamma_U$, and the final row corresponds to lengths of time agents spend in the $I$ state, which does correspond to an exponential distribution with mean $\gamma_I$. The first column uses a time unit of 1 day, the second column uses a time unit half a day and the last column uses a time unit of a quarter of a day.}
    \label{fig:correction}
\end{figure}

Whilst conducting this investigation we noticed that the agent-based approach overestimated the result, as can be seen in Figure \ref{fig:no_correction_conv}, but converges towards the SEIR-D approach as we reduce $\Delta t$. When checking the rate parameters, we noticed that the observed mean was approximately $0.5 \Delta t$ greater than the expected mean from the fitted parameters, which can be seen in Figure \ref{fig:no_correction} both by value (in the title of each plot) and by the fact that the red line (the expected probability density function) mostly goes through the next histogram column at the corner. We also noticed that this was independent of the number of Monte Carlo realisations. This led us to add on a correction term to the fitted parameters to reduce the observed mean towards the expected mean, namely by solving
\begin{align*}
    \frac{1}{\gamma + x} = \frac{1}{\gamma} - \frac{\Delta t}{2},
\end{align*}
which rearranges to \eqref{correction_term}. We note that $c(\gamma;\Delta t) \rightarrow 0$ as $\Delta t \rightarrow 0$. Utilising this correction term, we see that the observed means for the parameters are significantly more accurate in Figure \ref{fig:correction} and the output of the agent-based approach matches the SEIR-D approach better in Figure \ref{fig:correction_conv}. It is important to stress that whilst we do see convergence as $\dt$ tends to 0, significantly more computational power is needed when reducing $\dt$. By adding the correction term, we can use a larger time step, and thus less computational power, whilst still maintaining accurate results. This is particularly important when one could consider using a agent-based approach to model a complex phenomena, say one which does not have an obvious equation-based approach, and still be able to conduct parameter estimation. 

As for the probability parameters, we noticed that as the number of Monte Carlo realisations increased, the closer the probabilities got to the SEIR-D parameter, which can be seen in Table \ref{tab:prob_params}. Intuitively, this is to be expected as the more decisions being made, the closer the probability should be approximated. One also notices that you can calculate the probabilities at the end of the simulation (rather than count) by manipulating the SEIR-D approach in the following way. To obtain $m_U$, one integrates \eqref{RUeq} and rearranges to find
\begin{align*}
    \gamma_U \int_0^T U(s) \, \ds = \frac{1}{1 - m_U}(R_U(T) - R_U(0)),
\end{align*}
which, by integrating \eqref{DUeq}, inserting above and rearranging, gives
\begin{align*}
    m_U = \frac{D_U(T) - D_U(0)}{R_U(T) - R_U(0) + D_U(T) - D_U(0)}.
\end{align*}
One can apply the same idea to get
\begin{align*}
    m_H = \frac{D_H(T) - D_H(0)}{R_H(T) - R_H(0) + D_H(T) - D_H(0)},
\end{align*}
and
\begin{align*}
    p = \frac{U(T) - U(0) + \gamma_U \int_0^T U(s) \, \ds}{U(T) - U(0) + \gamma_U \int_0^T U(s) \, \ds + I(T) - U(0) + \gamma_I \int_0^T I(s) \, \ds}.
\end{align*}

\begin{table}[!htb]
    \centering
    \caption{Observed values of the probability parameters using a different number of Monte Carlo iterations to compare the observed value against the expected value given the state and fitted parameter, where $m_H$ is the proportion of individuals who die in hospital (calculated by $\mu_H (\gamma_H + \mu_H)^{-1}$).}
    \label{tab:prob_params}
    \begin{tabular}{lccccc}
        \toprule
         & \multicolumn{4}{c}{Monte Carlo iterations} & \\
        \cmidrule{2-5}
        Parameter & 1 & 5 & 10 & 20 & Expected value \\
        \midrule 
        $p$ & 0.9440 & 0.9411 & 0.9424 & 0.9405 & 0.9401 \\ 
        $m_U$ & 0.0019 & 0.0017 & 0.0015 & 0.0014 & 0.0013 \\ 
        $m_H$ & 0.4480 & 0.3573 & 0.3935 & 0.3815 & 0.3843 \\
        \botrule
    \end{tabular}
    
\end{table}

\end{document}